\documentclass{aa}

\usepackage{graphicx}
\usepackage{epsfig}

\begin{document}

   \title{A New Solution to the Plasma Starved Event Horizon Magnetosphere}

   \subtitle{Application to the Forked Jet in M87}

   \author{B. Punsly
          \inst{1}
          \
          M. Hardcastle\inst{2}
          \and
          K.Hada \inst{3}}

   \institute{1415 Granvia Altamira, Palos Verdes Estates CA, USA
90274: ICRANet, Piazza della Repubblica 10 Pescara 65100, Italy and
ICRA, Physics Department, University La Sapienza, Roma,
Italy\\
\email{brian.punsly@cox.net}
         \and
             Centre for Astrophysics Research, School of
Physics, Astronomy and Mathematics, University of Hertfordshire,\\
College Lane, Hatfield AL10 9AB, UK\\\
\and
 Mizusawa VLBI Observatory, National Astronomical Observatory of
Japan, Osawa, Mitaka, Tokyo 181-8588, Japan\\
             }

   \date{Received October 15, 2017; }
\titlerunning{Plasma Starved Event Horizon Magnetospheres}

\abstract {Very Long Baseline Interferometry observations at 86 GHz
reveal an almost hollow jet in M87 with a forked morphology. The
detailed analysis presented here indicates that the spectral
luminosity of the central spine of the jet in M87 is a few percent
of that of the surrounding hollow jet $200 -400 \mu\rm{as}$ from the
central black hole. Furthermore, recent jet models indicate that a
hollow ``tubular" jet can explain a wide range of plausible
broadband spectra originating from jetted plasma located within
$\sim 30\mu\rm{as}$ of the central black hole, including the 230 GHz
correlated flux detected by the Event Horizon Telescope. Most
importantly, these hollow jets from the inner accretion flow have an
intrinsic power capable of energizing the global jet out to
kiloparsec scales. Thus motivated, this paper considers new models
of the event horizon magnetosphere (EHM) in low luminosity accretion
systems. Contrary to some models, the spine is not an invisible
powerful jet. It is an intrinsically weak jet. In the new EHM
solution, the accreted poloidal magnetic flux is weak and the
background photon field is weak. It is shown how this accretion
scenario naturally results in the dissipation of the accreted
poloidal magnetic flux in the EHM not the accumulation of poloidal
flux required for a powerful jet. The new solution indicates less
large scale poloidal magnetic flux (and jet power) in the EHM than
in the surrounding accretion flow and cannot support significant EHM
driven jets. }


   \maketitle
%

\section{Introduction}
Large scale poloidal magnetic flux that threads the event horizon
(EH) of a rotating black hole (BH) forms an event horizon
magnetosphere (EHM) that is a viable source of BH driven jets. Since
the BH cannot be a source of plasma, the EHM is charge starved
(lacks a supply of charge that is required to support a frozen-in
magnetosphere, everywhere) and strong analogies with pulsar driven
winds have been made \citep{bla77}. A large distinction between
these two environments is that the neutron star (NS) is a
superconductor and supports magnetic fields with the largest field
strengths in the known Universe ($\sim 10^{8}\rm{G} -
10^{14}\rm{G}$), whereas the BH cannot support its own magnetic
field, since the field must be produced outside of the EH
\citep{pun08}. This distinction has not been considered in depth in
previous treatments of EHM jets. This paper discusses plausible
astrophysical circumstances in which this distinction has a crucial
effect on the physics of the system.

\par The EHM is located within the vortex of the accretion flow and requires
plasma injection in order to maintain a jetted system \citep{bla77}.
In seminal efforts, two viable options for producing the plasma in
the EHM were postulated. The first was drawn directly from pulsar
theory. In the charge starved limit, various types of vacuum gaps
and null (zero density) surfaces can exist in principle. As in
pulsar theory, the semi-vacuum electric field in these gaps can
accelerate leptons to very high energy thereby powering multi-stage
pair creation scenarios that seed the magnetosphere with an ideal
magnetohydrodynamic (MHD) plasma \citep{stu71,che86}. Analogously,
EHM gap models always assume that a background magnetic field is
already present \citep{bes92,hir98,hir16,bro15,pti16,lev00,lev11}.
The other idea, unique to BHs, is that the ambient $\gamma$-ray
field (presumably from the accretion flow) can produce enough
electron-positron pairs to seed the EHM. This study considers these
scenarios in the context of creating (as opposed to perpetuating) an
EHM in realistic astrophysical environments. In particular, in any
BH time evolution problem a causal temporal order of events is
required to establish the initial state. Without this key element as
part of the solution, it is not clear that a physical solution is
attained.
\par In Section 2, the details of the
scenario in which the EHM is created by a slow accumulation of thin,
weak, isolated magnetic flux tubes that are transported to the EH by
an accretion flow is explored. This assumed model of the seeding of
the EHM is the basis of the analysis of the charge starved limit
discussed in this paper. By ``thin, weak, isolated magnetic flux
tubes" it is meant:
\begin{itemize}
\item thin: the dimensions of the flux tubes are very narrow compared to the
dimensions of the disk and BH,
\item isolated: the large scale
poloidal flux that extends above or below the disk consists of a few
strands of flux that extend off to infinity as opposed to closing as
loops back into the disk. They accrete sporadically as opposed to a
nearly uniform, continual deposition of flux tubes into the EHM
\item weak: the field lines are readily deformed by the
surrounding disk atmosphere. The field strength is less than that
which is required to initiate a pair cascade.
\end{itemize}
The EHM solution considered here is evaluated in the charge starved
limit. Without sufficient plasma, it is shown that the accreted
poloidal magnetic flux readily dissipates in the EHM. The
dissipation is rapid relative to the rate that plausible accretion
scenarios can replenish the flux. Thus, a highly magnetized EHM is
not created.

\par The radio galaxy, M87, appears
to be an ideal candidate for the new EHM solution. It has a very low
luminosity accretion flow with arguably too low a photon flux to
support significant pair creation on weak accreting flux tubes in
the EHM. Furthermore, new high resolution Very Long Baseline (VLBI)
86 GHz VLBI observations resolve the jet in M87 on scales much
closer to the central BH than has been accomplished for any other
radio loud active galactic nucleus (AGN) \citep{kim16,had16}. These
images reveal a jet with an unexpected forked topology that seems to
represent a hollow jet (see Section 4). There is no evidence of
significant jet emission along the central spine above the event
horizon in agreement with the new EHM solution to be presented in
this paper. The new EHM solution is particularly relevant in the
context of recent models of hollow jets emanating from the inner
regions of an accretion flow that can describe a very wide range of
plausible broadband spectra (mm wavelengths to UV) of the base of
the jet in M87 on scales $\sim 15 -30\mu\rm{as}$. In addition to
explaining broadband emission from the region that produces the
correlated 230 GHz flux detected by the Event Horizon Telescope
(EHT), the jet base has sufficient power to energize the entire jet
out to kiloparsec scales \citep{pun18}. There is no need to invoke a
powerful invisible spine jet driven by the EHM in order to power the
jet \citep{mos16}. This supports the most direct interpretation of
the 86 GHz VLBI images: the jet is hollow because the EHM jet is
intrinsically weak in accord with the model presented here. Thus
motivated, much of the discussion to follow is focused on the
example of M87.

The paper is organized as follows. Section 2 is a discussion of the
details of the time evolution of weak, isolated flux tubes in the
completely charge starved limit. This section assumes negligible
plasma injection into the EHM in order to describe the new solution
of the EHM that is proposed here. Without plasma injection from the
external environment or a particle creation gap in the weak flux
tubes, currents cannot be maintained. Flux is dissipated, not
accumulated, if it accretes to the EHM. In Appendices A -C, the
details of the dissipation of the poloidal magnetic flux transported
within the charged starved, accreting flux tube is explored by means
of approximate solutions to Maxwell's equations in curved spacetime.
The lack of a reservoir for accreted flux in the EHM indicates a
weak EH driven jet.
\par The second part of the paper focuses on the application of the model to M87. The new model of the
magnetosphere is predicated on the mode of accretion and inefficient
pair creation. It is shown in Appendix D that for any plausible
model there is some minimum field strength below which the posited
accreted flux tubes will not produce a potential difference across
the vacuum gap large enough to initiate a pair cascade. Thus, pair
production in an external $\gamma$ -ray field would be required to
seed the EHM with plasma and would determine the maximum sustainable
magnetic field and jet power in the EHM. In Section 3, the
observational evidence that bounds the $\gamma$ -ray luminosity of
the inner accretion flow, from above, in M87 is discussed. No
$\gamma$ -ray telescope can resolve the inner accretion flow. The
highest resolution observations of the hard photon spectrum are with
the Chandra X-ray telescope. The core flux within 0.67 arcsec of the
nucleus is extracted. This is combined with broadband hard photon
spectra of active galactic nuclei from INTEGRAL in order to give
bounds on the $\gamma$ -ray luminosity from the nucleus. This in
turn implies an upper bound on the maximum sustainable magnetic
field strength in the EHM and the resultant maximum Poynting flux
that can be delivered by an EHM jet in M87. It is concluded that M87
is likely an example of a source with a weak $\gamma$-ray field near
the EH that is incapable of producing enough pairs to support the
currents required for an astrophysically significant EHM. In Section
4, it is noted that the results of Sections 2 and 3 and Appendix C
indicate that M87 is a possible example in which the EHM is so
charge starved that any jet produced in this region will be very
weak. It is shown that HSA (High Sensitivity Array) observations at
86 GHz support the new EHM model. There is a profound nadir of
emissivity along the central spine at the jet base above the
putative EHM that is consistent with this basic consequence of the
new EHM solution. In the following, it is assumed that $ M = 6
\times 10^{9} M_{\odot}$ ($8.4 \times 10^{14} \rm{cm}$ in
geometrized units) appropriate for M87 \citep{geb11}.
\begin{figure*}
\includegraphics[width= 0.5\textwidth]{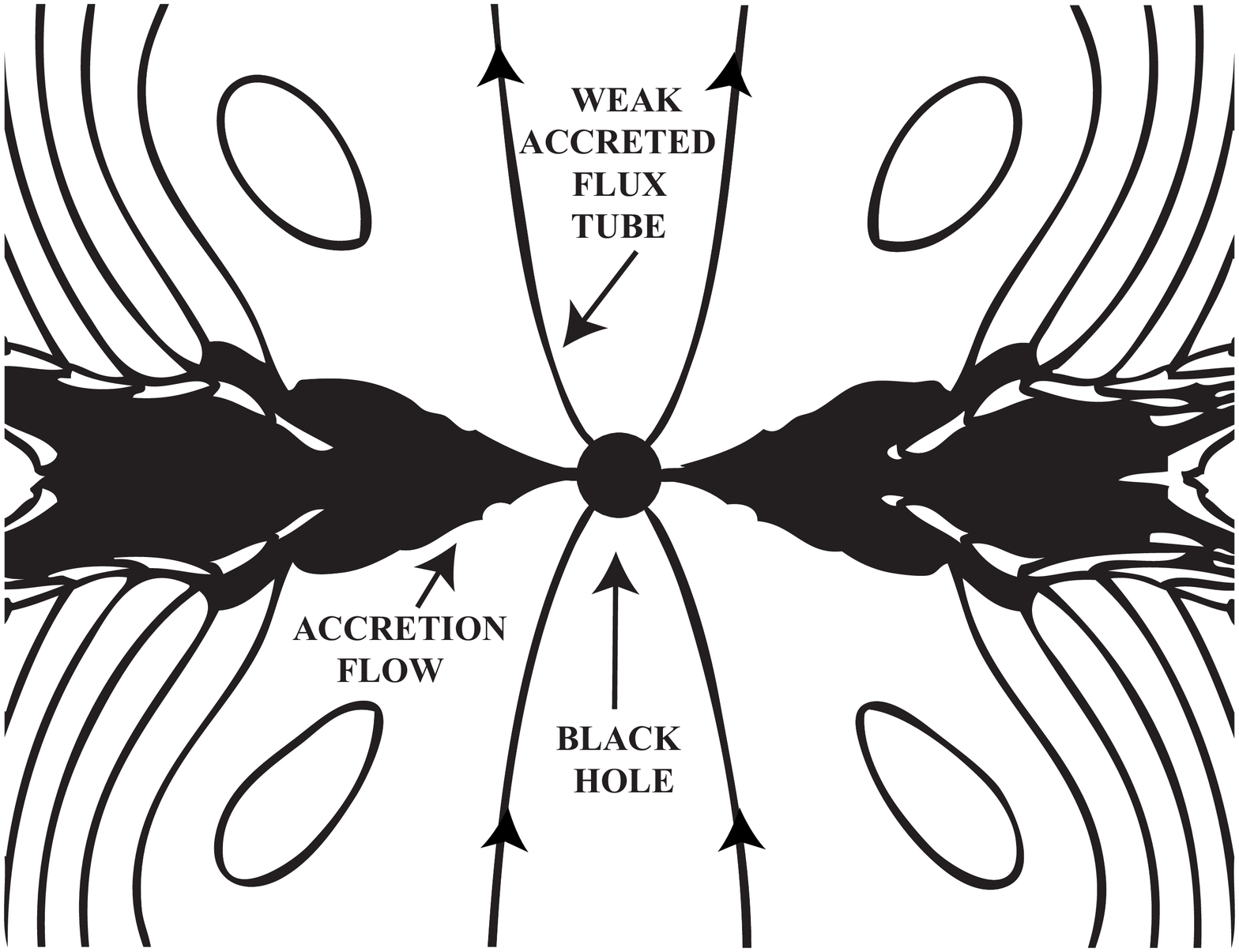}
\includegraphics[width= 0.5\textwidth]{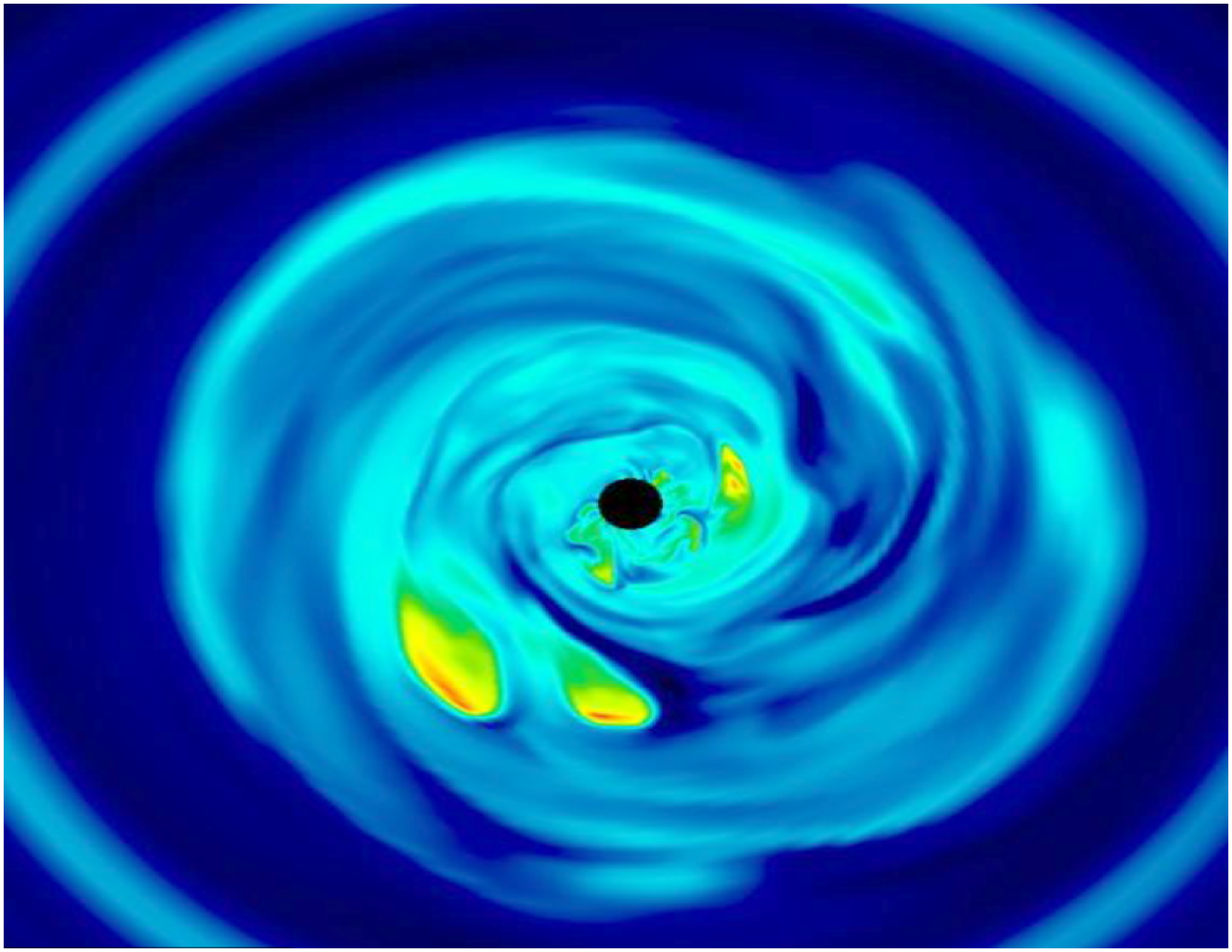}
\caption{The accretion of weak flux tubes into the EHM occurs in the
SFMHD+MF simulation of \citet{bec09} in the left hand frame. The
right hand frame shows isolated flux tubes in the 3-D radiatively
inefficient simulations of \citep{igu08} and \citep{pun09}. The
strength of the vertical poloidal magnetic field is color coded.
Dark blue is no field and red is a strong field (near equipartition
with the gas pressure). The inner calculational boundary is a circle
of radius 2M. Notice the weaker, green, small patches of vertical
flux in the inner accretion flow. See the text for more details.}
\end{figure*}
-------------------------------------------------------
\section{The Creation of an EHM by Accretion}
This study considers a possible new EHM solution that might occur in
some astrophysical black hole accretion systems. It is predicated on
a particular mode of accretion onto a rotating (Kerr) BH described
by a mass, $M$, and an angular momentum per unit mass, $a$. The
context is the initial seeding of the BH magnetosphere with large
scale poloidal flux. This is the initial state for the time of
evolution of the EHM. The specific details of how an EHM is
established are not known, and the processes involved are on too
small a scale to be observed directly, even if one were to be
observing during the initial stages. Thus any scheme for
establishing the EHM must rely on assumptions. It is known that the
flux must be delivered from the external environment since the Kerr
BH does not support a magnetic field in isolation.  A plausible
method of creating a significant EHM is the radiatively inefficient
accretion of weak poloidal magnetic flux from large distances over a
long period of time \citep{igu08,bec09,mck12}. Similar ideas have
been proposed for protostellar systems \citep{ler99}. It has been
suggested that the large scale magnetic flux near a black hole has
its origins in the advection of the weak large scale patchy magnetic
field in the intergalactic medium or from a magnetized stellar wind
or a tidal disruption event of a nearby magnetized star
\citep{mck12}. This is the scenario considered in this model of the
EHM and it is the fundamental assumption of this paper.

\par Note that the charge starved limit and the assumed pair creation in an EHM violates
perfect MHD. There are no existing numerical simulations that can
study this limit. Perfect MHD numerical experiments involving
accreting mass always end with the code crashing before the
charge-starved limit is approached (see \citet{mei01} and references
therein). Thus, numerical simulations artificially insert a non-MHD
mass floor that perpetuates the solution \citep{dev03,mck04}. One
can distinguish these from ideal MHD simulations by denoting them as
SFMHD+MF (single fluid MHD plus mass floor) in the following. A
numerical simulation that utilizes a mass floor is not an acceptable
device if one is considering the time evolution of a charge depleted
system as is the case here. Thus, the dynamics of the charge starved
accretion into the EHM will be described in what follows by
approximate analytic arguments.

The dynamics take place in the background spacetime of a rotating
black hole, the Kerr solution. In Boyer--Lindquist coordinates, the
Kerr metric, $g_{\mu\nu}$, is given by the line element in
geometrized units

\begin{eqnarray}
&&\mathrm{d}s^{2} \equiv g_{\mu\nu}\, \mathrm{d}x^{\mu}\mathrm{d}
x^{\nu}= -\left (1-\frac{2Mr}{\rho^{2}}\right)
\mathrm{d}t^{2}+\rho^{2}\mathrm{d}\theta^{2} \nonumber \\
&& +\left(\frac{\rho^{2}}{\Delta}\right)\mathrm{d}r^{2}
-\frac{4Mra}{\rho^{2}}\sin^{2}\theta\, \mathrm{d}\phi \,
\mathrm{d}t  \nonumber \\
&& +\left [(r^{2}+a^{2})+\frac{2Mra^{2}}{\rho^{2}}\sin^{2}
\theta\right ] \sin^{2} \theta \, \mathrm{d}\phi^{2} \; ,
\end{eqnarray}
where $\rho^{2} = r^{2}+a^{2}\cos^{2}\theta$ and
\begin{equation}
\Delta = r^{2}-2Mr+a^{2} \equiv \left(r-r_{{+}})(r-r_{{-}}\right ).
\end{equation}

There are two event horizons given by the roots of the equation
$\Delta=0$. The outer horizon at $r_{{+}}$ is of physical interest
\begin{equation}
r_{{+}}=M+\sqrt{M^{2}-a^{2}} \; .
\end{equation}
In order to simplify the calculations, one can compute quantities in
a hypersurface orthogonal, orthonormal frame. There exists an
orthonormal, Zero Angular Momentum Observers (ZAMO) frame associated
with each coordinate pair, $(r,\, \theta)$. The ZAMOs can be used to
express, locally, the electromagnetic field in terms of electric and
magnetic (observer-dependent) fields. There are three main benefits
of calculating in the ZAMO frames. The orthonormality condition is
beneficial for utilizing many results and techniques from special
relativity. By contrast, the Boyer-Lindquist coordinates are
curvilinear and not even orthogonal. Thus, a physical interpretation
of the covariant and contravariant quantities near the black hole is
far from trivial. Secondly, unlike other orthonormal frames, being
hypersurface orthogonal, the ZAMO frame provide an unambiguous
definition of the electromagnetic field that is integrable
\citep{pun08}. Most importantly, as shown in Appendix B, one can
rotate the poloidal direction to always be along the local poloidal
magnetic field direction. This greatly simplifies the interpretation
of the electromagnetic quantities. Even though calculations are much
clearer in the rotated ZAMO basis, ultimately we need to express the
results in terms of the Boyer-Lindquist coordinates associated with
the stationary observers at asymptotic infinity. Thus, we describe
the transformation between frames. The ZAMO basis vectors are
\begin{eqnarray}
&& \hat{e}_0 = \alpha_{\rm{Z}} ^{-1}\left(\frac{\partial}{\partial t} + \Omega_{\rm{Z}}\frac{\partial}{\partial\phi}\right) \;\nonumber \\
&& \Omega_{\rm{Z}} =\frac{-g_{\phi\, t}}{g_{\phi\phi}} \;,\; \alpha_{\rm{Z}} =\frac{\sqrt{\Delta} \sin{\theta}}{\sqrt{g_{\phi\phi}}}\;, \nonumber \\
&&\hat{e}_\phi=\frac{1}{\sqrt{g_{\phi\phi}}}\frac{\partial}{\partial\phi}\
,\;\hat{e}_r = \left( \frac{\Delta^{1/2}}{\rho} \right)
\frac{\partial}{\partial r} \ , \;\hat{e}_\theta =\left(
\frac{1}{\rho} \right) \frac{\partial}{\partial \theta} \; .
\end{eqnarray}
The lapse function, $\alpha_{\rm{Z}}$, is the gravitational redshift
of the ZAMOs as measured by the stationary observers at asymptotic
infinity (i.e., astronomers on earth). Note that
\begin{eqnarray}
&&\lim_{r \rightarrow \infty} \alpha_{\rm{Z}} = +1 \; ,\\
&&\lim_{r \rightarrow r_{+}} \alpha_{\rm{Z}} = 0 \;.
\end{eqnarray}
Similarly, $\Omega_{\rm{Z}}$, is the angular velocity of the ZAMOs
as viewed by stationary observers at asymptotic infinity.

The basis covectors are

\begin{eqnarray}
&&\omega^{\hat 0}=\alpha_{\rm{Z}}dt\; ,\; \omega^{\hat
r}=\sqrt{g_{rr}}dr\; , \nonumber \\ && \omega^{\hat
\theta}=\sqrt{g_{\theta\theta}}d\theta\;, \;\omega^{\hat
\phi}=\sqrt{g_{\phi\phi}}d\phi \;.
\end{eqnarray}
Boyer-Lindquist evaluated quantities are distinguished from ZAMO
evaluated quantities by the use of a "tilde" on the variables. Both
formalisms will be utilized in the description of the flux
evolution.
\subsection{Weak Isolated Flux Tubes in the EHM} The concept of a weak isolated magnetic flux tube is
introduced by means of SFMHD+MF simulations. In the initial state
there is no large scale poloidal flux that threads the event
horizon. There needs to be a mechanism that can transport large
scale poloidal flux to the EHM. The accretion flow is the natural
place to look for such a source. Attempts to spontaneously create
the flux from the accretion flow itself by means of the
magneto-rotational instability (MRI) proved to be unsuccessful
\citep{bec09}. A simulation requires a net poloidal flux in the
accretion flow in order to build up a significant EHM
\citep{igu08,bec09}. When the simulation starts there is a transient
state when the first flux tubes approach the event horizon. It will
look similar to the t=1500M snapshot from a SFMHD+MF simulation of
\citet{bec09}, depicted in Figure 1. The magnetic flux is clearly
weaker in the EHM than in the disk and a single field line is
separated by a large gap from the magnetic field in the disk. This
is an accreted isolated flux tube created in the early stages of a
SFMHD+MF simulation. All transient early stages of SFMHD+MF
simulations create an EHM by beginning with the arrival of a first
weak flux tube, unless the initial state is unphysical and posits
large amounts of flux in the initial state in the EHM proper or
adjacent to the EHM. This is true even if a saturated magnetosphere
is attained at large times \citep{ mck12}.
\par The right hand frame of Figure 1 shows a different depiction of
isolated flux tubes in the 3-D radiatively inefficient simulation of
\citep{igu08} and \citep{pun09}. This frame is from the on-line
movies of the latter reference. The strength of the vertical
poloidal magnetic field is color coded. Dark blue is no field and
red is a strong field (near equipartition with the gas pressure of
the surrounding accretion flow). Notice that the field accumulates
in isolated patches. Even though it was definitely not the intent of
this simulation, in this image there are small patches of weak field
near the inner boundary (a circle of radius 2M). The greenish-yellow
patches have a magnetic pressure $\sim 2\% -10\%$ of equipartition
with the gas pressure of the surrounding gas. These are examples of
weak isolated flux tubes. It is important to note that in this
simulation they formed as a consequence of the amalgamation of a
steady influx of very weak field from the outer calculation
boundary. The flux reservoir at the outer boundary is axisymmetric,
but the accretion flow is not. The 3-D accretion flow is driven by
the MRI as in the \citet{bec09} simulations. However, these
simulations have a much larger reservoir of flux at the outer
boundary. If there is a large reservoir of poloidal flux,
condensations of vertical flux will naturally occur as a consequence
of the MRI driven turbulence. In general the isolated flux tubes are
more magnetized in other time snapshots. However, these simulations
suggest that weak isolated vertical flux tubes might be natural in
an accretion flow. The patches of vertical magnetic flux near the
black hole should be weaker and more isolated if the reservoir of
flux is a weak patchy intergalactic magnetic field as opposed to a
constant flood of flux as in the simulation in Figure 1.
\subsection{Relevant Assumptions of SFMHD+MF Simulations} In this
paper, the early time behavior of a nascent EHM is analyzed after
abandoning some major assumptions of the SFMHD+MF simulations. In
particular:

\begin{enumerate}
\item The notion of a mass floor is dropped. Physically, this
equates to a black hole accretion system in which there is no
efficient plasma injection mechanism to support the flux in the EHM.
\item There is no large
reservoir of magnetic flux that persistently deposits flux into the
EHM. It is instead assumed that the flux deposits into the EHM on
astronomically large time scales. For example, the jet propagation
speeds indicate a jet lifetime of $>10^{6}$ years for many radio
loud AGN \citep{wil99}. This is $>10^{8}$ light travel times across
the black hole in M87. Even a small fraction of this time scale is
not computer resource efficient for SFMHD+MF simulations, so a more
compact flux source is assumed in those numerical models. However, a
compact source is not a valid assumption if the rate that flux
accretes is dynamically important as in this section and Appendix C.
\item It is also not assumed that the distant flux reservoir is uniform,
but is composed of small distinct patches of isolated flux,
\end{enumerate}
By dropping assumption 1), there will be insufficient plasma to
support MHD. In the low or zero pair creation limit, it is shown in
Appendix C that the magnetic flux will dissipate in the EHM on a
timescale, $t_{\rm{dis}}$. that is estimated. Dropping assumption 3)
allows for an non uniform deposition of flux into the EHM over time.
This naturally produces temporal gaps between episodes in which
isolated patches of accreted flux are deposited in the EHM. The
dynamical timescale to deposit more flux, $t_{\rm{dyn}}$, can exceed
$t_{\rm{dis}}$ allowing the flux tube to dissipate before an
accumulation of flux can occur.
\subsection{Maxwell's Equations Description of a Weak Isolated flux
Tube} Figure 2 shows an idealized isolated, large scale, poloidal
flux tube accretion scenario. There are two components of the
magnetic field in the accretion flow since the system is in rotation
with the plasma, $B^{\phi}$ and $B^{P}$, azimuthal and poloidal
respectively.  In the thin flux tube limit (so thin that cross-field
gradients in the current and field are negligible compared to the
gradients at the boundaries), the electromagnetic sources are
approximately surface currents. To quantify this for flux tubes that
emanate from the accretion disk, a cylindrical coordinate system in
flat space is chosen for demonstrative purposes,
$(\rho^{\rm{cyl}},\, \phi, \, z)$. The inner boundary of the flux
tube is $\rho^{\rm{cyl}}_{-}(z)$ and the outer boundary is
$\rho^{\rm{cyl}}_{+}(z)$, where axisymmetry is assumed for
simplicity. The thin flux tube limit is defined for small
$\epsilon>0$ by the conditions,
\begin{eqnarray}
&&\frac{\mid B^{\phi}(\rho^{\rm{cyl}}_{-}(z)-\epsilon, \,z)\mid }{\mid B^{\phi}(\rho^{\rm{cyl}}_{-}(z), \,z)\mid}\ll 1\\
&&\frac{B^{\phi}(\rho^{\rm{cyl}}_{+}(z)+\epsilon, \,z)\mid }{\mid B^{\phi}(\rho^{\rm{cyl}}_{+}(z), \,z)\mid}\ll 1\\
&&\frac{\mid B^{P}(\rho^{\rm{cyl}}_{-}(z)-\epsilon, \,z)\mid }{\mid B^{P}(\rho^{\rm{cyl}}_{-}(z), \,z)\mid}\ll 1\\
&&\frac{\mid B^{P}(\rho^{\rm{cyl}}_{+}(z)+\epsilon, \,z)\mid }{\mid B^{P}(\rho^{\rm{cyl}}_{+}(z), \,z)\mid}\ll 1\\
&&\frac{\mid B^{\phi}(\rho^{\rm{cyl}}_{-}(z), \,z) -
\mid B^{\phi}(\rho^{\rm{cyl}}_{+}(z), \,z)\mid }{\mid B^{\phi}(\rho^{\rm{cyl}}_{-}(z), \,z)\mid}\ll 1\\
&&\frac{\mid B^{P}(\rho^{\rm{cyl}}_{-}(z), \,z) - \mid
B^{P}(\rho^{\rm{cyl}}_{+}(z), \,z)\mid }{\mid
B^{P}(\rho^{\rm{cyl}}_{-}(z), \,z)\mid}\ll 1 \\
&&\frac{\rho^{\rm{cyl}}_{+}(z)-\rho^{\rm{cyl}}_{-}(z)}{\rho^{\rm{cyl}}_{-}(z)}\ll
1
\end{eqnarray}

The fact that the slowly accreting isolated flux tubes have a $B$
field much stronger than that of the plasma on both sides of the
flux tube means that the surface current will change the field
strength from approximately zero to $B$ in Ampere's law at the inner
face of the flux tube. Similarly, the surface current will change
the field from $B$ to near zero at the outer face of the flux tube.
Since the flux tube accretes with the plasma in the disk, it
essentially spirals with the Keplerian velocity with a relatively
slow inward radial drift \citep{sad11}. Thus, to first
approximation, one can ignore displacement current in Amperes's law
for the field inside the axisymmetric flux tube. Let $\mathbf{K}$
designate a surface current. In the approximately cylindrical
configuration. by Ampere's Law and Equations (8) - (14),
\begin{eqnarray}
&&\frac{4\pi}{c}K^{\phi}(\rho^{\rm{cyl}}_{-}(z), \,z) \approx -B^{P}(\rho^{\rm{cyl}}_{-}(z), \, z)\\
&&\frac{4\pi}{c}K^{\phi}(\rho^{\rm{cyl}}_{+}(z), \,z) \approx B^{P}(\rho^{\rm{cyl}}_{+}(z), \, z)\\
&&\frac{4\pi}{c}K^{P}(\rho^{\rm{cyl}}_{-}(z), \,z) \approx \rho^{\rm{cyl}}_{-}(z)B^{\phi}(\rho^{\rm{cyl}}_{-}(z), \, z)\\
&&\frac{4\pi}{c}K^{P}(\rho^{\rm{cyl}}_{+}(z), \,z) \approx -\rho^{\rm{cyl}}_{+}(z)B^{\phi}(\rho^{\rm{cyl}}_{+}(z), \, z)\\
&&\frac{4\pi}{c}K^{\phi}(\rho^{\rm{cyl}}_{-}(z), \,z) \approx - K^{\phi}(\rho^{\rm{cyl}}_{+}(z), \,z)\\
&&\frac{4\pi}{c}K^{P}(\rho^{\rm{cyl}}_{-}(z), \,z) \approx -
K^{P}(\rho^{\rm{cyl}}_{+}(z), \,z)
\end{eqnarray}
The integral of $K^{P}$ over an orthogonal cross-section of either
the inner or outer boundary of the flux tube (the total poloidal
current) is approximately conserved from the disk to asymptotic
infinity in the axisymmetric, magnetically dominated limit and a
conserved value represents electromagnetic angular momentum flux
conservation in the flux tube \citep{pun08}. $K^{\phi}$ is set by
the poloidal magnetic flux conservation condition from the accretion
flow to asymptotic infinity in each flux tube. The corresponding
curved spacetime versions of these surface current equations are
derived in Appendix B in the ZAMO frames.
\begin{eqnarray}
&&\frac{4\pi}{c}K^{\phi}_{\rm{Z}}(r_{\rm{in}}, \, \theta_{\rm{in}}) \approx -B^{P}(r_{\rm{in}}, \, \theta_{\rm{in}})\\
&&\frac{4\pi}{c}K^{\phi}_{\rm{Z}}(r_{\rm{out}}, \,
\theta_{\rm{out}}) \approx B^{P}(r_{\rm{out}}, \,
\theta_{\rm{out}})\\
&&\frac{4\pi}{c}K^{P}_{\rm{Z}}(r_{\rm{in}}, \, \theta_{\rm{in}}) \approx B^{\phi}(r_{\rm{in}}, \, \theta_{\rm{in}})\\
&&\frac{4\pi}{c}K^{P}_{\rm{Z}}(r_{\rm{out}}, \, \theta_{\rm{out}})
\approx -B^{\phi}(r_{\rm{out}}, \, \theta_{\rm{out}})\;.
\end{eqnarray}
These equations are required near the black hole. The
Boyer-Lindquist coordinates, $(r_{\rm{in}}, \, \theta_{\rm{in}})$
indicates a point on the inner boundary of the flux tube and
$(r_{\rm{out}}, \, \theta_{\rm{out}})$ indicates a point on the
outer boundary of the flux tube.
\subsection{The Dynamics of Accreted Weak Isolated Flux
Tubes} During the inflow through the disk, the source of the charges
that create the currents that sustain the magnetic flux is in the
base of the flux tube that is frozen into the accretion flow. Plasma
is shot outward by magneto-centrifugal forces in the rotating flux
tube and dragged inward near the base (accretion) by gravity
\citep{igu08}. The plasma that is shot outward is provided by the
accretion flow before the flux tube enters the EHM.\begin{figure}
\includegraphics[width= 0.5\textwidth]{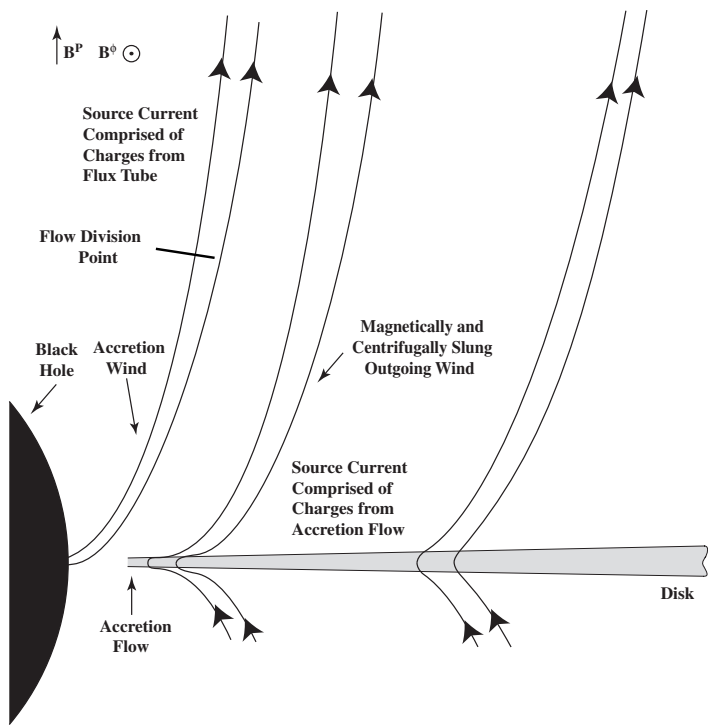}
\caption{ The dynamics of the accretion of weak flux into the EHM is
depicted above. The accretion flow in this image is likely much
thinner than what occurs in M87. The concepts that are illustrated
are not a consequence of the accretion flow thickness.}
\end{figure}

\par Figure 2 shows that the dynamic that existed in the flux tube in the accretion disk
persists as it enters the EHM. In particular, plasma is still shot
outward by magneto-centrifugal forces in the rotating flux tube and
still dragged inward (accretion) by gravity \citep{mei01,sem04}. At
the flow division surface, the flow divides into an accretion flow
and an outgoing wind \citep{phi83}. Due to gravitational redshifting
and frame dragging, the plasma in the flux tube near the EH is out
of causal contact with the large scale poloidal flux \citep{pun08}.
In this discussion, it is assumed that there is no external plasma
injection mechanism such as pair production (see Appendix D and
Section 3 for the likelihood of this possibility in M87). Plasma
that is already threaded on the flux tube must provide the outgoing
plasma and the currents supporting the magnetic field.
\footnote{Note that there is no dynamic at the disk-EHM boundary
that naturally changes the MHD flux tube, with its local current
system, into a flux tube in which the source is transferred to a
surface current that resides at the inner surface of the disk.}.
There is a finite amount of plasma in the flux tube and the plasma
quickly becomes tenuous. The plasma starts to drain from the flow
division surface producing a vacuum gap as depicted in Figure 3. The
figure is a schematic diagram that shows the split that occurs in
the distribution of plasma, not the field lines, as the vacuum gap
begins to expand. Initially, the field lines are not severed in the
vacuum gap. However, the poloidal magnetic field is not uniform in
this region. The poloidal field bulges outward and inward as
fringing effects become pronounced. A laboratory example of this
effect would occur if one split a long solenoid in the middle, then
pulled the two halves away from each other along the axis of
symmetry. A non-uniform bulging field occurs in the gap between the
two coils. At later times, the fringing fields associated with the
spreading vacuum gap expand and can approach other fringing field
lines along circles (due to axisymmetry) of X-type reconnection
points. This reconnection process can change the topology of the
poloidal magnetic field.
\begin{figure}
\includegraphics[width= 0.5\textwidth]{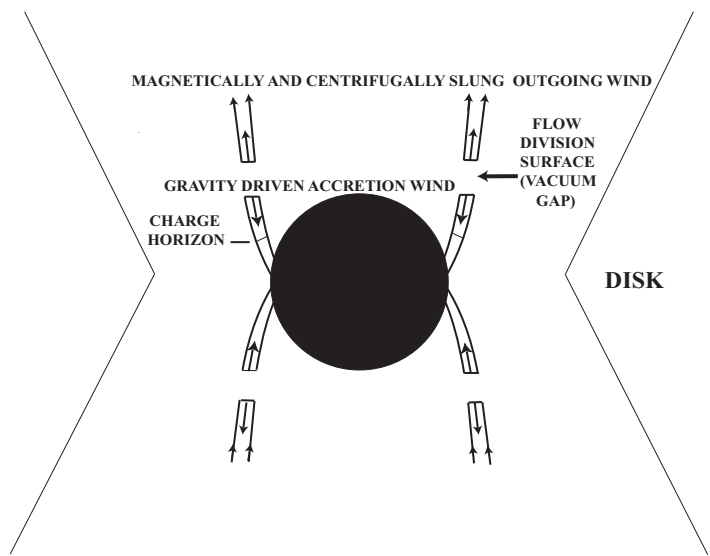}
\caption{ In the charge-starved limit, a vacuum gap will spread
outward  from the flow division surface if there is no substantive
pair injection mechanism as is quite possibly the case in M87. In
this charge starved limit, the surrounding magnetosphere is a
semi-vacuum. The only significant source of electromagnetic fields
is within the flux tube. The very tenuous stray charges have
trajectories that are affected by these fields, but the number
density is too small to provide a source for a significant
perturbation to these fields. Charges can only flow inward across
the charge horizon \citep{kom03,pun04}}
\end{figure}
\par In Appendix C, we discuss a model of an accreted flux tube in
which there are insufficient charges to maintain the source currents
in the EHM - charge starved. The lower portion of the flux tube
contracts toward the black hole by gravity and the outer is slung
out by magneto-centrifugal forces. Evaluating Maxwell's equations as
the inner portion of the flux tube approaches the event horizon
indicates that the large scale poloidal magnetic field in the EHM
will approximately be a decaying magnetic dipole (see Figure C.2).
Since the calculation is very long and involved, we only summarize
the logic and results in the main text.
\begin{enumerate}
\item Equations (21) -(24) are used to describe the current
distribution in the ingoing portion of the severed flux tube as two
nested, coaxial helical surface currents distributions, one in each
hemisphere.
\item In Section C.1, it is shown that due to gravitational redshift
as these helical current flows accrete close to the event horizon
they seem to be frozen in corotation with the horizon, hovering just
above it, as viewed by external observers. Thus, these axisymmetric
electromagnetic source are approximately time stationary to the
external observers (eg. in Boyer-Lindquist coordinates) that would
be affected by the large scale poloidal magnetic field. Therefore,
Laplace's equations can be used to accurately depict the large scale
poloidal magnetic field for these sources at any given
Boyer-Lindquist time, $t$ \citep{pun89}.
\item In Section C.2, it is shown that at late times in the accretion, near the event
horizon, the large scale poloidal magnetic field from the helical
current sources can be approximated as the large scale poloidal
magnetic field due to 4 azimuthal current rings that are located
near the black hole (see Figure C.1).
\item In Section C.3, the large scale poloidal magnetic field produced by the four
current loops is calculated by means of Laplace's equations in
curved spacetime and the results plotted in Figure C.2. The large
scale poloidal magnetic field is approximately a decaying magnetic
dipole.
\item In Section C.4, it is estimated that the flux tube dissipates (magnetic dipole
decays) on a time scale, $t<10M$, after the vacuum gap starts to
spread apart. This time scale is much less than any time scale of
the accretion flow. Thus, for the accretion scenario posited in this
section, the flux will dissipate before more flux can accumulate in
the EHM. A highly magnetized EHM will not form.
\item In Section C.4,  based on Figure C.2, it is argued
that surface currents induced in the disk during the field decay do
not prevent accreted, thin, isolated flux tubes in a charge starved
EHM from dissociating. These currents are decaying and are of the
wrong sign to maintain the accreted flux within the EHM.
\end{enumerate}

This suggests that an interesting new dynamic can exist in the EHM
if the EHM is charge starved: no vacuum gap pair cascades and weak
$\gamma$-ray pair production. Thus motivated, standard vacuum gap
pair production in the EHM are considered in Appendix D and
$\gamma$-ray pair production is discussed in the case of M87 in the
next section.

\section{The $\gamma$-Ray Induced Pair Creation in the EHM of M87}

\begin{figure*}
\begin{center}
\includegraphics[width= 0.8\textwidth]{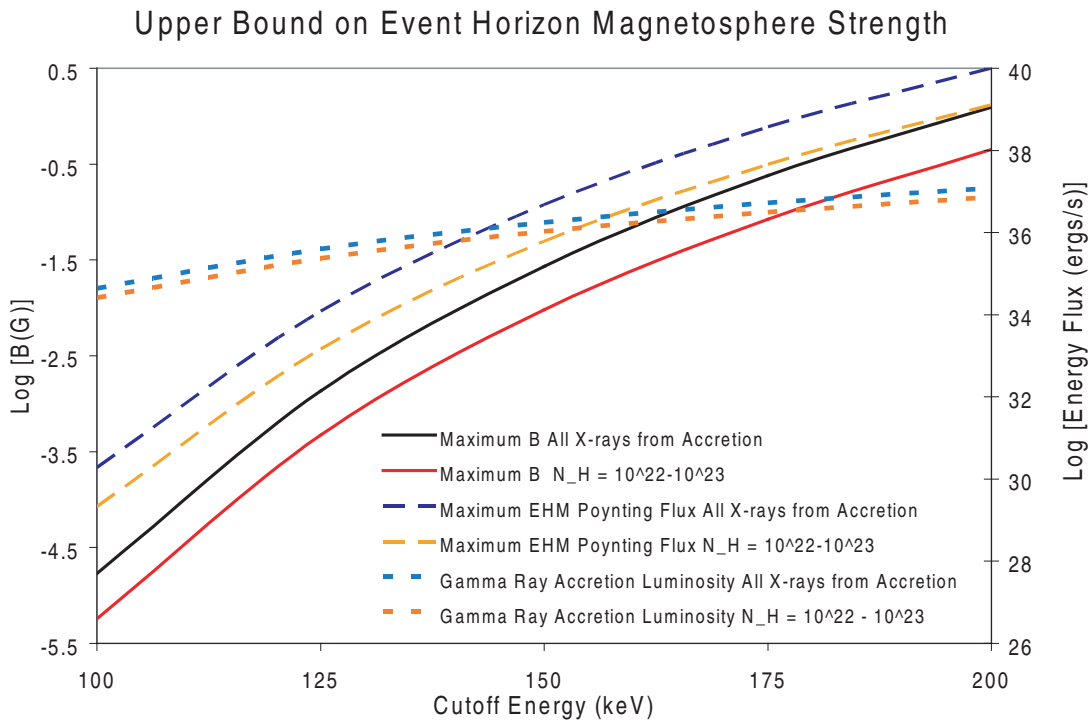}

\caption{Plots of the upper limits on $L_{C}$ and the maximum
sustainable Poynting flux from the EHM for M87 as a function of
$E_{c}$ for two cases, $L_{x}$ is entirely from accretion or $L_{x}$
from accretion is attenuated by an absorbing screen, $N_{H}$. The
maximum sustainable magnetic field in the EHM assuming that
$a/M=0.9$ is also plotted.}
\end{center}
\end{figure*}
In the weak field limit, proposed in the last section, the EHM will
not be able to sustain pair creation in a vacuum gap (see Appendix D
for more elaboration). Thus, pair creation in an external
$\gamma$-ray field is required in order to provide plasma to the
accreted flux tubes and this will determine the maximum sustainable
magnetic field strength in the EHM. This particle injection
mechanism is considered in the context of the accretion scenario of
Section 2 in the environment of M87.
\par $\gamma$-rays from the jet in M87 are produced relatively far away and beamed away
from the EH and do not contribute to EHM pair production. However,
the $\gamma$-ray field of the accretion flow can produce
electron-positron pairs in the EHM. In this section, the available
data related to the hard photon spectrum of M87 is considered in
order to make as precise as possible any constraints that can be
imposed on the $\gamma$-ray luminosity. The resolution of telescopes
in the $\gamma$-ray band is many orders of magnitude too low to be
of any use. However, the low energy region of the hard photon
spectrum can be resolved to within 0.67 arcsec by Chandra. This
information is used in consort with what is known about the hard
photon spectra of other AGN (in particular, the cutoff energy) in
order to constrain the $\gamma$-ray luminosity in M87. Even though
it will be concluded that the Chandra flux is likely from the jet
itself, this detection still provides a useful and non-arbitrary
bound on the hard photon spectrum from the accretion flow.

\par The number density of created pairs from a background
$\gamma$-ray field can be estimated by balancing the infall
(free-fall) rate with the pair creation rate \citep{phi83}

\begin{equation}
n \sim \left( \frac{m_p}{m_e} \right) \left(
\frac{L_{_C}}{L_{_{Edd}}} \right)^2 10^{13} M^{-1}_{8} \,
\mbox{cm}^{-3} \; ,
\end{equation}

\noindent where $L_{_C}$ is the luminosity of $\gamma$-rays $> 1$
MeV from the accretion flow, $L_{_{Edd}}$ is the Eddington
luminosity and $M_{8}$ is the mass of the black hole in units of
$10^{8}M_{\odot}$. If the pair creation process can produce a charge
density in excess of the Goldreich-Julian density, $ \rho_{_{G-J}}$,
then the growth of the electric field in the vacuum gap can be
quenched and the surface current flow sustained on the flux tube
\citep{gol69}. One can estimate $ \rho_{_{G-J}}$ near the EH
\begin{eqnarray}
&& \rho_{_{G-J}}\sim \frac{\Omega_{F} B}{2 \pi ce} \sim \sigma
\frac{10}{M_{8}}\left(\frac{B}{10^{4}\,\mathrm{G}}
\right)\mathrm{cm}^{-3} \;, \\
&& \Omega_{F} \equiv \sigma \Omega_{H} \;,
\end{eqnarray}

\noindent where $\Omega_{F}$ and $\Omega_{H}$ are the angular
velocity of the magnetic field and the event horizon angular
velocity as viewed from asymptotic infinity, respectively. For a
given $\gamma$-ray field, the condition, $n_{e}>\rho_{_{G-J}}$,
determines the maximum sustainable $B$ field in a thin accreting
magnetic flux tube in the EHM.

\par $L_{_C}$ in M87 is constrained by revisiting the estimate of the accretion flow X-ray
luminosity, $L_{x}$, from \citet{har09}, with a smaller extraction
region (correcting for the PSF outside the region) of 0.67 arcsec
(versus 1 arcsec) to avoid contamination from the knot, HST-1, in
the Chandra data \citep{har03}. No detectable X-ray excess above a
single unabsorbed power law flux density was observed: $\alpha_{x} =
1.1$, $L_{E} \propto E^{-\alpha_{x}}$, where $E$ is photon energy
and $L_{x}= 2.9 \times 10^{40}\rm{erg/s}$ from  2-10 keV. The
nucleus is a continuation of the large scale X-ray jet with similar
values of $L_{x}$ and $\alpha_{x}$ to those of the knots in the jet
\citep{wil02}. Mid-IR and optical studies conclude that there is no
hidden strong accretion source, but just a synchrotron nuclear
source in M87 \citep{why04,chi99}. Broadband correlations amongst
the nuclear synchrotron and X-ray fluxes in many Fanaroff-Riley I
(FRI) radio galaxies such as M87 also imply a jet origin for X-rays
\citep{har00,har09}.
\par An upper bound for $L_{_C}$ due to accretion can be estimated in two ways from the Chandra data.
First, consider the limiting scenario (although it is unlikely
considering the discussion above) that the Chandra nuclear flux is
from the accretion flow. This estimate is performed in order to
establish the most conservative limit on the upper bound on
$L_{_C}$, Secondly, it is assumed that the accretion X-ray source is
hidden by an attenuating column of neutral hydrogen,
$10^{22}\rm{cm}^{-2}< N_{H}< 10^{23}\rm{cm}^{-2}$ and $\alpha_{x} =
0.7$ \citep{har09}. Note that there is no evidence of such a large
$N_{H}$ in M87. In this case, an intrinsic $L_{x}< 1.9 \times
10^{39}\rm{erg/s}$ from 2-10 keV with 90\% confidence is estimated.
These are ``worst case," not necessarily likely, scenarios for
producing upper bounds on $L_{_C}$.
\par The wideband $L_{x}(\rm{wb})$ from accretion in AGN and Galactic compact objects
is typically approximated by a cutoff power law, $L_{x}(\rm{wb})
\propto E^{-\alpha_{x}} \rm{e}^{-E/E_{c}}$, where $E_{c}$ is the
cutoff energy \citep{mal14}. It is assumed that the spectral index,
$\alpha_{x}$, is constant from keV to MeV energies in the following
calculations. However, the upper bounds that are computed below are
valid as long as the power law does not flatten at higher energies.
Figure 4 contains plots of three upper bounds as functions of
$E_{c}$ for both scenarios: $L_{_C}$, the associated maximum
sustainable Poynting flux from the EHM and the maximum sustainable
value of $B$ from Equations (25)-(27). The $B$ plot assumes the
seminal value of $\sigma=0.5$ from \citet{bla77} and $a/M=0.9$. The
range of $E_{c}$ appropriate to the putative accretion source of
$L_{x}(\rm{wb})$ is motivated by INTEGRAL observations indicating an
average $E_{c} =125$ keV for type I AGN and radio loud AGN in which
$L_{x}(\rm{wb})$ is not of blazar (jet) origin \citep{mal14}. The
MHD Poynting flux in the magnetically dominated limit is
\begin{equation}
\int{S^{P}\mathrm{d}A_{_{\perp}}} =
k\frac{\Omega_{F}^{2}\Phi^{2}}{2\pi^{2} c} \approx
\frac{\Omega_{F}^{2} (4\pi B(r_{+}^2))^{2}}{2\pi^{2} c}\;,
\end{equation}
where $\Phi$ is the total magnetic flux enclosed within the jet
(through the EH), $\mathrm{d}A_{_{\perp}}$ is the cross-sectional
area element (surface area element of EH) and $k$ is a geometrical
factor that equals 1 for a uniform highly collimated jet
\citep{pun08}. Using the fact that $\Omega_{H}= a/(2Mr_{+})$ and
Equations (25) - (28), the upper bound on the approximate Poynting
flux is independent of BH spin and the jet model for $\sigma$ over a
wide range: $0.4< a < 0.95$ and $0.1< \sigma <1$.
\par Figure 4 shows that the Chandra data likely imply a $\gamma$-ray accretion source in M87 that is
insufficient to support even a 1G field in a charge-starved EHM.
Furthermore, the largest upper bounds on Poynting flux are more than
three to four orders of magnitude less than the estimated jet power
of $\sim 10^{43} \rm{ergs/s} - 10^{44} \rm{ergs/s}$.
\citep{mcn11,sta06}.

\begin{figure*}
\begin{center}

\includegraphics[width= 0.70\textwidth]{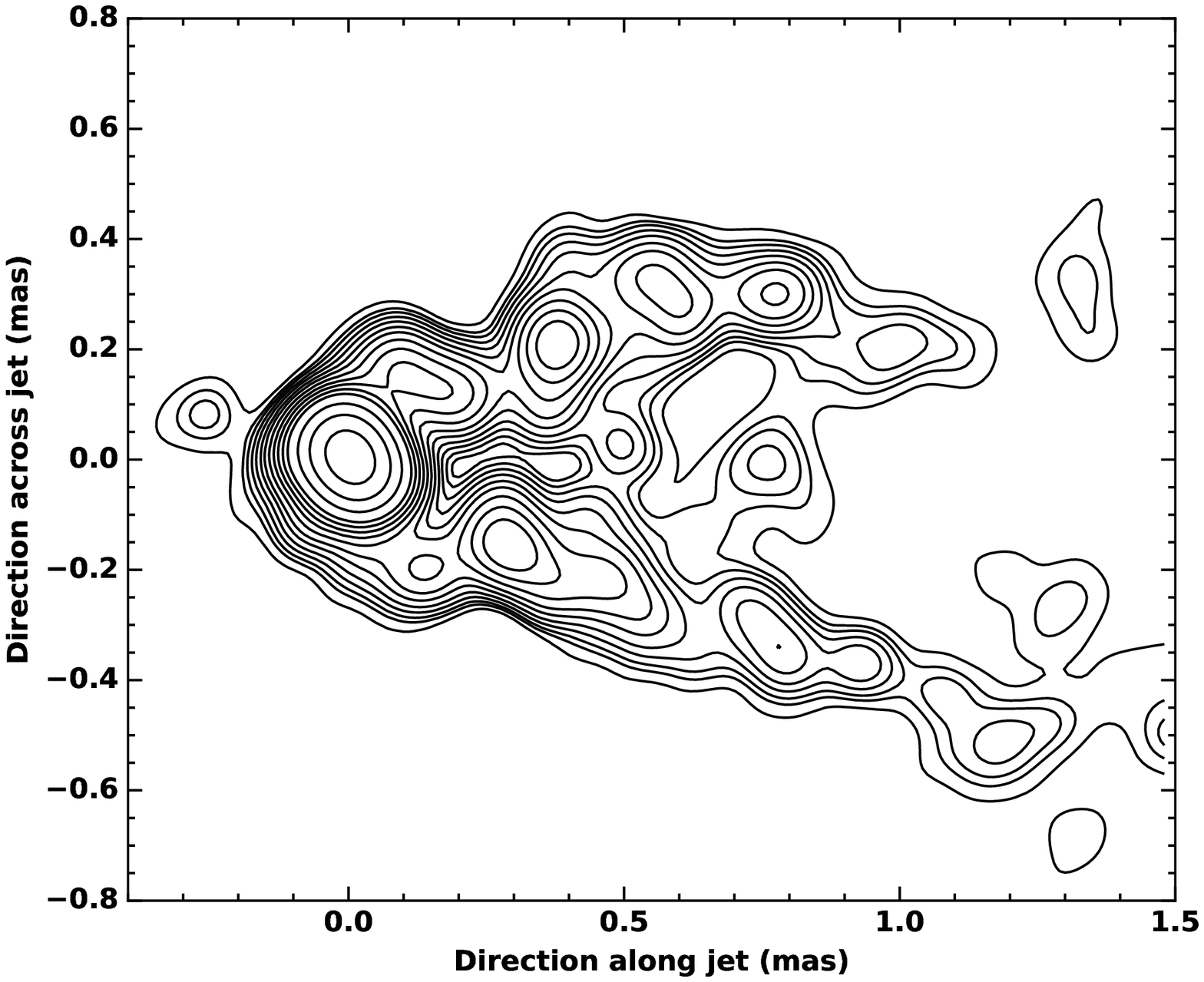}
\includegraphics[width= 0.70\textwidth]{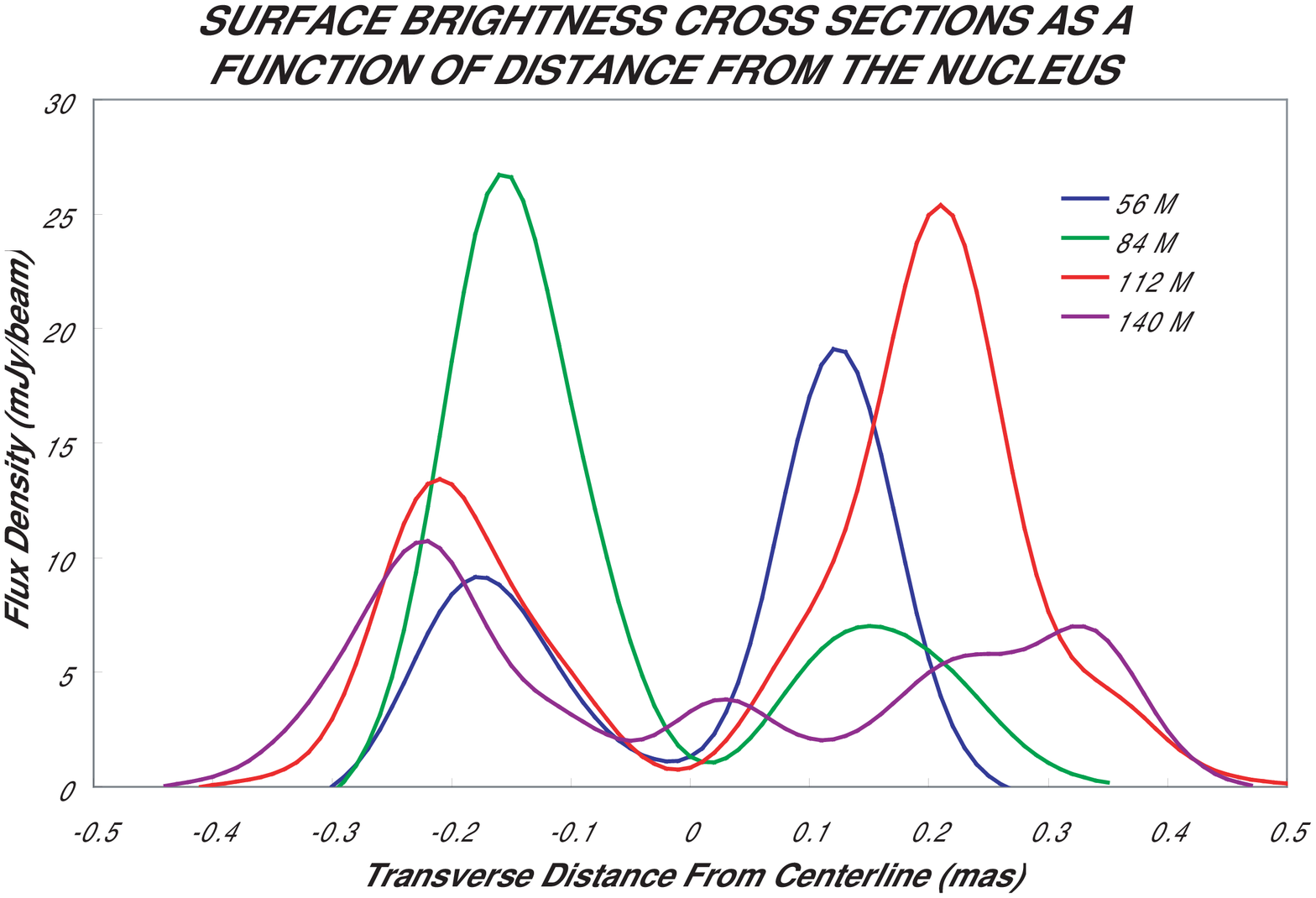}

\caption{The central flux nadir of the jet near its base is apparent
in the 86 GHz HSA image from Hada et al. (2016)  (restored with a
0".0001 beam) in the top frame. The bottom frame plots surface
brightness cross-sections from the image above. The central flux
nadir is resolved within 56 M (0".0002) of the EH. The central flux
nadir surface brightness is $\sim 4\% - 8\%$ of the average surface
brightness on the limbs, 56 M - 112 M from the EH and $\sim 20 -
30\%$ of the average surface brightness of the limbs at 140 M.}
\end{center}
\end{figure*}

\section{Evidence for a Hollow Jet in M87}

\par This section considers possible evidence in support of
the posited model of the EHM for the particular case of M87. New
data reductions from high sensitivity 86 GHz VLBI are provided that
indicate a much larger deficit of luminosity along the jet spine at
the base of the jet in M87 than has been previously demonstrated at
lower resolution. This result is combined with lower resolution data
in order to examine the details of the new EHM model and previous
explanations of limb brightening.
\par  The fundamental testable consequence of this model of the EHM is
the existence of a jet with a base that is wider than the EH (i.e.,
driven from the accretion flow) that will have a dearth of intrinsic
emissivity along its central spine, above the EH. The jet in M87 is
likely optically thin since the flux density, $F_{\nu} \propto
\nu^{-0.8}$ \citep{had16}. If the jet is hollow to first order,
lines of sight (LOS) that are nearly parallel (or anti--parallel) to
the tangent to the circumference of the jet will intersect larger
column densities of optically thin plasma than a LOS through the
middle of the jet. Thus, one expects a limb brightened appearance in
two places, one where the LOS is parallel and one where it is
anti-parallel to the tangent of the jet circumference if the jet is
hollow to first order as predicted by the new EHM model. In order to
test this prediction, consider the HSA image at 86 GHz in Figure 5
\citep{had16}. The flux nadir along the center of the jet is
resolved within 56 M (0.2 mas) of the EH and is not transient,
occurring in multiple epochs \citep{kim16}. The flux nadir from 0.2
mas to 0.4 mas can be described quantitatively in terms of the integrated
flux density. The total 86 GHz flux density of the central flux
nadir in the region 0.2 mas - 0.4 mas from the black hole is
$\approx 6\%$ of the flux density of the surrounding outer sheath
jet (hollow jet).

\par Patches of enhanced surface brightness are clearly detected in
the central void (``the spine") at 0.5 mas in Figure 5 and beyond 1.5
mas in multiple epochs with increasing prominence downstream
\citep{mer16}. In particular, the velocity field of the inner jet in
M87 has been mapped by means of a 43 GHz VLBA wavelet analysis
\citep{mer16}. Even though, the data is from 2007, seven years before
the HSA observations, the components line up reasonably well with the
ridges seen in the 86 GHz image between 0.5 mas and 1.5 mas. The 43
GHz wavelet analysis is consistent with new 22 GHz VLBI data from 2014
\citep{had17}. Within 1.5 mas, the wavelet based apparent velocities
are similar to the values obtained by \citet{had16} for the HSA
observation, $\sim~0.1c -0.4 c$, quite subluminal. The 43-GHz analysis
also provides valuable evidence of the dynamics of the spine beyond
1.5 mas from the core. The apparent velocity, $v_{\rm{app}}$, of the
individual components of the spine, at the smallest displacements from
the core for which the signal to noise of the spine is sufficient for
such estimates (1.5 mas - 2 mas from the core), is $v_{\rm{app}/}/c =
1.33 \pm 0.63 $ and $v_{\rm{app}/}/c = 1.16 \pm 0.77 $ for the
surrounding limbs \citep{mer16}. The similarity of the velocity field
for the spine and the limbs suggests that the spine is gradually being
filled by plasma that originates in the surrounding sheath and slowly
spreads inward towards the central axis, as would be expected in the
model in which the spine is empty at the jet base. In other words, the
$v_{\rm{app}}$ distribution and increased spine prominence downstream
is well explained in terms of a weak EHM jet surrounded by a hollow
jet that slowly fills in the relative void with kinematically similar
plasma as it propagates.

We consider a few possible alternative models for the observations.

\subsection{Bifurcating Obstacle} There could be an obstacle $\leq 120M$ downstream
from the black hole. When the jet collides with this obstacle, it
would bifurcate, rendering the central parts of the jet empty
without it being intrinsically so. However, the jet has the hollow
morphology in multiple epochs \citep{kim16}. So there needs to be a
quasi-stationary feature hovering $\leq 120M$ above the black hole.
We know of no physical mechanism that could create such a
quasi-equilibrium above the black hole.

\subsection{Doppler Suppression} The central spine could be
of similar emissivity to the observed sheath, but have a much higher
speed, so that Doppler suppression reduces the observed spine
surface brightness. Given the Doppler factor for the approaching jet
${\cal D} = 1/(\Gamma[1-\beta\cos\theta)]$, where $\Gamma$ is the
bulk Lorentz factor, it can easily be shown that Doppler suppression
takes place for angles to the line of sight $\theta >
\cos^{-1}[(\Gamma
  -1)/(\Gamma^2-1)^{1/2}]$. For example, the bulk Lorentz factors
($\Gamma \sim 10$--$50$) often implied by observations of
superluminal motion in blazars, \citet{lis16}, Doppler suppression
will take place unless the angle to the line of sight is smaller
than a critical angle in the range $\sim 25^\circ - \sim 10^\circ$.
While this model cannot be ruled out in principle, we regard the
observed similar apparent velocities in the sheath and spine region
as evidence against it; in such a model we might expect to see
higher apparent speeds in the center of the jet.

\subsection{Ghost Jet} {The central spine could have the same speed as the
observed sheath, but have a low emissivity because the energy
density of the particles is low \citep{mos16}, forming a `ghost
jet'.} However, it is not obvious that the Poynting flux core can be
protected from an infusion of high energy particles, if it is
surrounded by an energetic outflow of protonic material from the
surrounding disk/corona accretion system. There are three
significant sources of high energy particles.

\par First, the accretion vortex in numerical simulations of
radiatively inefficient accreting systems is not the ordered
force-free environment envisioned in theoretical treatments when the
putative Poynting jet does exist in the EHM. In the simulations of
\citet{kro05}, it was found that the EHM and the jet base are very
unsteady and the accretion vortex appeared to be a cauldron of
strong MHD waves rather than what would be expected of a force-free
structure (even though the energy density of the particles is much
less than the energy density of the electromagnetic field). This
appears to be the case in the simulations of \citet{tch12}, as well,
based on the supporting online movies in which the field lines in
the vortex whip around chaotically. As these strong MHD waves crash
against the bounding sheath jet, fast magnetosonic shocks are
created. Even though, in this magnetically dominated limit, these
shocks are not highly effective at accelerating plasma to high
energy (see \citet{ken84}), there would be many such shocks. This
would be expected to imbue the Poynting flux core with a back-flow
of particles from the high energy tail of the plasma that is
energized at the shock front.

\par Secondly, it is difficult to keep the sheath plasma from mixing
into the jet, if it is there. Near the base of the jet, it was found
in 3-D numerical simulations that the corona/jet interface is
unsteady with large fingers of hot gas being injected into the
Poynting jet on scales of $\sim 20M - 30 M$ from the BH
\citep{pun07}. To accurately model such mixing of corona and jet gas
requires an accurate numerical scheme. For example, codes like HARM
which is used in \citet{tch12,mck12} do not utilize the contact
discontinuity in their Riemann solver. The absence of the contact
discontinuity tends to numerically dissipates effects associated
with abrupt density gradients \citep{pun16}. Furthermore, a recent
study of \citet{how17} showed that the typical numerical resistivity
in MHD simulations is large enough that mixing modes such as the
Kelvin-Helmholtz instability (associated with a strong magnetic
coronal loop) are highly suppressed. Thus, it is an open question
how much the corona and sheath will seed a putative strong Poynting
jet core with high energy plasma.

\par Thirdly, the chaotic behavior in the accretion vortex and the
large toroidal twisting of the field lines is not conducive to
maintaining an ordered, untangled field. Field tangling is often
called braiding in solar physics. Braided fields are believed to
release the extra energy of tangling as they relax to a more
simplified state by reconnection \citep{wil10}. Reconnection of the
braided fields in the jet can also provide high energy plasma to the
jet and the fields are strongest near its base \citep{wil10,bla15}.
\par Based on the fact that the putative ghost jet would support a pair
cascade of high energy particles in the accretion vortex,
\citet{bro15}, and the three plausible sources of high energy plasma
described above, it is not at all clear that the energy density of
the jet can be maintained low enough to keep it invisible or
extremely weak at mm wavelengths. Thus, the study of alternative
scenarios that require fewer assumptions, such as a weak EHM jet, is
worthwhile.

\section{Conclusion} This paper considers an EHM that is built up by the
accumulation of accreted weak, isolated strands of magnetic flux
over a long period of time. In the absence of a significant
background photon field, an analysis based on Maxwell's equations in
curved spacetime that was developed in Appendices A - C indicates
that the magnetic flux will readily dissipate in the EHM instead of
accumulate in the EHM. In this accretion scenario, the resultant
weak field that can be sustained in the EHM is determined by the
pair creation rate in the $\gamma$ -ray field of the accretion
disk/corona. In Section 3, evidence that M87 appears to have a weak
$\gamma$ -ray accretion source was presented based on the Chandra
X-ray spectrum of the nucleus and the high energy cutoffs of other
AGN derived form INTEGRAL observations. The derived upper bounds on
the $\gamma$ -ray luminosity renders the EHM of M87 ineffectual for
jet launching. In Section 4, it is shown that 86 GHz HSA
observations reveal a bizarre forked jet 50M - 400M from the black
hole. This is a manifestation of the weak central spine of the jet
above the EH that is expected as a consequence of the new solution
of the EHM. Many other FRI and some FRII radio galaxies also appear
to have weak accretion X-ray emission and likely weak $\gamma$ -ray
emission as well \citep{har09}. Thus, a weak or absent EHM might be
common to radio galaxies with radiatively inefficient accretion such
as M87. It is tempting to speculate that jet bases with a forked
morphology might occur in other radiatively inefficient radio
galaxies.
\par The EHM solution is consistent with recent hollow jet models from
the inner accretion flow of M87 \citep{pun18}. The models are able
to fit an extremely wide range of plausible spectra of broadband
emission emanating from $15-30 \mu\rm{as}$ scales including the 230
GHz correlated flux detected by the EHT. For high spin black holes,
$a/M=0.99$, the jet transports $10^{43} -10^{44} \rm{ergs/sec}$ if
the poloidal magnetic field is 8 - 15 G in the inner accretion flow.
Thus, these models can supply the entire jet power of M87 that has
been estimated from the analysis of large scale features
\citep{mcn11,sta06}. The accord with constraints based on broadband
spectra and jet power is achieved with a magnetic field strength
that is consistent with assumption 1) of Section 2. In particular,
based on Appendix D, $ 8 - 15\rm{G}$ is $ \ll$ than the $ \sim 225
\rm{G}$ that would be required for a self-sustaining pair creation
mechanism on an accreted flux tube in the EHM in the absence of a
significant ambient soft photon flux. Thus, the key assumption of
the EHM solution presented here, a weak accreted magnetic field, is
a property of a wide range of high spin BH, hollow jets models of
M87 that have both a plausible mm wavelength to UV spectrum and a
jet power of $10^{43} -10^{44} \rm{ergs/sec}$.
\par The EHM solution described in this article could be used to
argue that a steady accretion of weak axisymmetric flux would also
dissipate in a charge starved EHM. But, more importantly, the flux
dissipation does not depend on the assumption of axisymmetry. Even
for non-axisymmetric flux tubes, as in the right hand frame of
Figure 1, the charges will drain off without a plasma source in the
EHM and the flux will be dissipated. Even though an axisymmetric
disk was used in the models of the broadband luminosity of the jet
in \citet{pun18}, this is not necessary to drive the jet from the
inner accretion flow. In the quasar jet launching study of
\citet{pun14}, the jets are considered to originate in isolated flux
tubes (magnetic islands), as in the right hand frame of Figure 1,
within the innermost accretion flow. In this case, the jet Poynting
flux is altered slightly from our Equation (28). Instead of the jet
power from the inner disk scaling as $(B^{P})^{2}$ as in Equation
(28), it scales as $(fB^{P})^{2}$, where $f$ is the filling fraction
of the disk threaded by isolated flux tubes with a vertical field
strength, $B^{P}$. It should be noted that in general (more
realistically) there would be a bivariate distribution of field
strengths and filling fractions. In the example of M87, as noted
above, for $a/M=0.99$ the broadband spectrum and jet power was fit
in \citet{pun18} with an inner accretion disk field strength of 8 -
15 G. For a filling factor, $f \sim 50\%$, this corresponds to $
B^{P} \sim 15 -30 \rm{G}$ in order to reproduce the jet power.

\par The EHM solution described in this article provides an alternative to assuming a
powerful invisible (or highly under-luminous) ghost jet along the
central spine on sub-mas scales that is also posited to be the
primary power source for the large scale jet on kpc scales. Being
under-luminous, by assumption, a powerful jet cannot be directly
verified by any observation on sub-mas or mas scales. It can only be
ascertained indirectly with deductive reasoning or it must dissipate
violently farther out in the jet, thereby revealing its intrinsic
power. Evidence of this second alternative, would be a spine that
far out shines the limbs over an extended region. Putative spine
emission on larger scales falls far short of satisfying this
requirement \citep{had18}. The heretofore only posited deductive
argument is that a powerful spine is required to energize regions of
enhanced emission such as the knot HST-1 nearly 1 arcsec from the BH
\citep{sta06,mer16}. However, in this context, it was shown in
\citet{pun18} that a hollow jet from the inner accretion flow not
only explains a multitude of plausible spectra of broadband emission
emanating from $15-30 \mu\rm{as}$ scales, but also supports $\sim
10^{44} \rm{ergs/s}$ of jet power. Thus, a powerful ghost jet is not
required to power the large scale jet (including energizing the knot
HST-1). This renders deductive arguments that the ghost jet must be
powerful in order to meet global energy requirements untenable. In
summary, a powerful ghost jet is not indicated directly by any
observation nor is it required to explain any of the observations.
\par By contrast, there are two very extreme properties in M87 that are observed near
the nucleus. Both are fundamental elements of the new EHM solution.
There is the extreme central flux nadir in the base of the jet near
the event horizon. There is also the extraordinarily weak high
energy luminosity of the accreting gas given the large central black
hole mass. The EHM solution presented here implies that these two
extreme circumstances might not be coincidental in M87. If the new
EHM solution applies to M87 then a luminous jet should extend back
towards its source in the inner accretion disk as in the hollow jet
models \citep{pun18}. The detection of a luminous forward jet on
scales $< 30 \mu\rm{as}$ by future EHT imaging would be direct
evidence of a powerful hollow jet connecting the accretion flow to
kpc scales and the compatible new EHM solution. This is in contrast
to models of ghost jets surrounded by a luminous sheath that predict
no strong forward jet emission at 230 GHz - 370 GHz on scales $< 40
\mu\rm{as}$ \citep{dex12,mos16}. Future EHT imaging might be able to
discriminate between these two models.

\begin{acknowledgements}
      We would like to thank Robert Antonucci for many valuable
      comments. This paper also benefitted from the insightful review of an anonymous
      referee.
\end{acknowledgements}

%
%

\begin{appendix} 
\section{Laplace's Equation in the Kerr Spacetime}
In \citet{pun89,pun08} and Appendix C, it is shown that the poloidal
magnetic field of axisymmetric electromagnetic sources near the
event horizon can be accurately described by Laplace's Equation as a
consequence of gravitational redshifting. Thus, Laplace's Equations
will be used in Appendix C.3 to compute the late time behavior of
the poloidal magnetic field of the spreading vacuum gap scenario
illustrated in Figure 3. This Appendix presents axisymmetric
solutions to Laplace's Equation.

\par In order to solve Laplace's Equation in the Kerr Spacetime, it is
customary to work with the spin coefficients of the field,
$\phi_{0}$, $\phi_{1}$, $\phi_{3}$, instead of the Faraday tensor,
$\tilde{F}^{\mu\nu}$ (tildes are used in the following to designate
Boyer-Lindquist evaluated quantities), since Maxwell's equations are
separable in the Newman-Penrose spin coefficients. One can
explicitly expand $\tilde{F}^{tr}$ in terms of the spin coefficients
\citet{pun08}:
\begin{eqnarray}
&& \tilde{F}^{tr} = \mbox{Re} \left\{ \frac{r^2+a^2}{\rho^2}
\phi_{_1} + \frac{\rm{i} a \tilde{\rho}^{*} \sin \theta}{\sqrt{2}}
\left( \phi_{_2} - \frac{\tilde{\rho}^2 \Delta \phi_{_0}}{2} \right)
\right\} \; ,\nonumber
\\[2pt]
&& \tilde{F}^{t\theta} = \mbox{Re} \left\{ \frac{\rm{i} a \sin
\theta}{\rho^2} \phi_{_1} - \frac{\left( r^2+a^2 \right)}{\sqrt{2}}
\frac{\tilde{\rho}}{\Delta} \left( \phi_{_2} - \frac{\tilde{\rho}^2
\Delta \phi_{_0}}{2} \right) \right\} \; ,\nonumber
\\[2pt]
&& \tilde{F}^{t\phi} = \mbox{Re} \left\{ - \frac{\rm{i} \tilde{\rho}
\rho^2}{2 \sqrt{2} \Delta \sin \theta} \left( \phi_{_2} +
\frac{\tilde{\rho}^2 \Delta \phi_{_0}}{2} \right) \right\} \;
,\nonumber
\\[2pt]
&& \tilde{F}^{r\theta} = \mbox{Re} \left\{ -
\frac{\tilde{\rho}^{*}}{2 \sqrt{2}} \left( \phi_{_2} +
\frac{\tilde{\rho}^2 \Delta \phi_{_0}}{2} \right) \right\} \;
,\nonumber
\\[2pt]
&& \tilde{F}^{r\phi} = \mbox{Re} \left\{ - \frac{a}{\rho^2}
\phi_{_1} - \frac{\rm{i} \tilde{\rho}^{*}}{\sqrt{2} \sin \theta}
\left( \phi_{_2} - \frac{\tilde{\rho}^2 \Delta \phi_{_0}}{2} \right)
\right\} \; ,\nonumber
\\[2pt]
&& \tilde{F}^{\theta\phi} = \mbox{Re} \left\{ - \frac{\rm{i}}{\rho^2
\sin \theta} \phi_{_1} + \frac{a \tilde{\rho}^{*}}{\sqrt{2} \Delta}
\left( \phi_{_2} - \frac{\tilde{\rho}^2 \Delta \phi_{_0}}{2} \right)
\right\} \; ,
\end{eqnarray}
\noindent where $\tilde{\rho}$ is given by
\begin{equation}
\tilde{\rho} = \frac{-1}{r - \rm{i} a \cos \theta} \; .
\end{equation}
Thus, knowledge of the spin coefficients is sufficient to determine
the electromagnetic field in Boyer-Lindquist coordinates. A
normalization change on the spin coefficients leads to simpler
solutions,
\begin{eqnarray}
&&
\Phi_0 = \phi_{_0} \; , \nonumber \\
&& \Phi_1 = \frac{\left(r - \rm{i} a \cos \theta \right)^2}
 {\left( r_{_+} - r_{_-} \right)^2} \phi_{_1} \; , \\
&& \Phi_2 = \frac{r - \rm{i} a \cos \theta}{\left( r_{_+} - r_{_-}
\right)^2} \phi_{_2} \; . \nonumber
\end{eqnarray}
Define $\,_{\pm 1} Y_{lm}$ as spin weighted spherical harmonics.
Also define,
\begin{eqnarray}
&& X \equiv \frac{ r - r_{_-} }{ r_{_+} - r_{_-} } \; , \quad Z_m
\equiv \frac{ma}{ r_{_+} - r_{_-} } \; .
\end{eqnarray}
The general solution to Laplace's equation in the Kerr space--time
for a source located between $r_1$ and $r_2$, with $r_{_+} < r_1 <
r_2 < \infty$ is presented \citep{bic76,pun08}.  In the region
between the source and the horizon, $r_{_+} < r < r_1$:
\begin{eqnarray}
&& \Phi_0 = \sum_{l,m} a_{lm} \, 2 [l(l+1)]^{-1} \left( 1 -
\frac{1}{X} \right)^{- \rm{i} Z_m} \nonumber\\
&&\times \frac{\rm{d}^2}{\rm{d} X^2} \left[ \,^2 y_{{lm}}^{(I)}
\right]
\,_{+1} Y_{lm} (\theta,\phi) \; ,\\[4pt]
\nonumber \\
&& \Phi_1 = \frac{\sqrt{2} \left( r_{_+} - r_{_-} \right)} { \left(
r - \rm{i} a \cos \theta \right)^2 }\nonumber\\
&& \times \sum_{l,m} a_{lm} \, [l(l+1)]^{-1} \left( 1 - \frac{1}{X}
\right)^{- \rm{i} Z_m}
 \left\{
 \left[l(l+1)\right]^{1/2}
 \right.
 \nonumber \\[4pt]
  &&
 \times
 \left[
 \left( r - \rm{i} a \cos \theta \right)
 \frac{\rm{d}}{\rm{d} X}
 \left( \,^2 y_{{lm}}^{(I)} \right) -
 \left( r_{_+} - r_{_-} \right)
 \left( \,^2 y_{{lm}}^{(I)} \right)
 \right]
 \,_0 Y_{lm} (\theta,\phi)
 \nonumber \\[4pt]
  &&
 \hspace{.1 in}
 \left.
 - \rm{i} a \sin \theta
 \frac{\rm{d}}{\rm{d} X}
 \left( \,^2 y_{{\ell m}}^{(I)} \right)
 \,_{+1} Y_{\ell m} (\theta,\phi)
 \right\}
+ \frac{E_a}{ \left( r - \rm{i} a \cos \theta \right)^2 } \; ,\\[4pt]
&& \Phi_2 = \frac{\left( r_{_+} - r_{_-} \right)^2} { \left( r -
\rm{i} a \cos \theta \right)^2 } \nonumber\\
&&\times \sum_{l,m} a_{lm} \left( 1 - \frac{1}{X} \right)^{- \rm{i}
Z_m}
 \left( \,^2 y_{{lm}}^{(I)} \right)
 \,_{-1} Y_{lm} (\theta,\phi) \; .
\end{eqnarray}
\noindent The solutions of most interest are those at $r
> r_2$.
\begin{eqnarray}
&& \Phi_0 = \sum_{l,m} b_{lm} \, 2 [l(l+1)]^{-1} \left( 1 -
\frac{1}{X} \right)^{- \rm{i} Z_m} \nonumber\\
&&\times \frac{\rm{d}^2}{\rm{d} X^2} \left[ \,^2 y_{{lm}}^{(II)}
\right] \,_{+1} Y_{lm} (\theta,\phi) \; ,
\nonumber \\[-6pt]
&&
\\
&& \Phi_1 = \frac{\sqrt{2} \left( r_{_+} - r_{_-} \right)} { \left(
r - \rm{i} a \cos \theta \right)^2 } \nonumber\\
&&\times \sum_{l,m} b_{lm} \, [l(l+1)]^{-1} \left( 1 - \frac{1}{X}
\right)^{- \rm{i} Z_m}
 \left\{
 \left[l(l+1)\right]^{1/2}
 \right.\nonumber \\[3pt]
 &&
  \hspace{-0.1 in}\times
 \left[
 \left( r - \rm{i} a \cos \theta \right)
 \frac{\rm{d}}{\rm{d} X}
 \left( \,^2 y_{{lm}}^{(II)} \right) -
 \left( r_{_+} - r_{_-} \right)
 \left( \,^2 y_{{lm}}^{(II)} \right)
 \right]
 \,_0 Y_{lm} (\theta,\phi)
 \nonumber \\[3pt]
 &&
 \left.
 - \rm{i} a \sin \theta
 \frac{\rm{d}}{\rm{d} X}
 \left( \,^2 y_{{\ell m}}^{(II)} \right)
 \,_{+1} Y_{\ell m} (\theta,\phi)
 \right\}
+ \frac{E_b}{ \left( r - \rm{i} a \cos \theta \right)^2 } \; ,
\\
&& \rule{0 in}{.1 in} \nonumber \\
&& \Phi_2 = \frac{\left( r_{_+} - r_{_-} \right)^2} { \left( r -
\rm{i} a \cos \theta \right)^2 }\nonumber\\
&& \times \sum_{l,m} b_{lm} \left( 1 - \frac{1}{X} \right)^{- \rm{i}
Z_m}
 \left( \,^2 y_{{lm}}^{(II)} \right)
 \,_{-1} Y_{lm} (\theta,\phi) \; .
\end{eqnarray}
\noindent The constants $a_{lm}$, $b_{lm}$, $E_a$, and $E_b$ are
determined by the nature of the source. The radial functions satisfy
\begin{equation}
\,^2 y_{{lm}}^{(I)} = \left( 1 - \frac{1}{X} \right)^{2 \rm{i} Z_m}
X(X-1) F(l+2, 1-l, 2-2 \rm{i} Z_m; X) \; ,
\end {equation}
\begin{equation}
\,^2 y_{{lm}}^{(II)} = \left( -X \right)^{-l} F(l, l+1-2 \rm{i} Z_m,
2l+2; X^{-1}) \; .
\end {equation}
The symbol, "$F$", stands for the hypergeometric function. Each
charge neutral solution is determined by the coefficients $a_{lm}$
and $b_{lm}$ which result from the spinorial current, $\,^2 J_{lm}
(\xi)$.

\begin{eqnarray}
&&\hspace*{-2pc} a_{lm} = - \frac{ 4 \pi (l+1)! \; \; \Gamma \left(
l + 1 - 2 \rm{i} Z_m \right)} { (2l+1)! \; \; \Gamma \left( 2 - 2
\rm{i} Z_m \right)}
\nonumber \\
&&\times \int_{X_1 - \epsilon}^{X_2 + \epsilon} \frac{
 \left( \,^2 J_{lm} (\xi) \right)
 \left( \,^2 R_{lm}^{(II)} (\xi) \right) }{ \xi (\xi-1) }
\rm{d} \xi \, ,\quad \nonumber \\
&&  \,^2 R_{lm}^{(II)} \equiv
\Bigg[1-\frac{1}{X}\Bigg]^{-iZ_m}\left[ \,^2 y_{{lm}}^{(II)} \right]
\, ,
\end{eqnarray}
\begin{eqnarray}
&&\hspace*{-2pc} b_{lm} = - \frac{ 4 \pi (l+1)! \; \; \Gamma \left(
l + 1 - 2 \rm{i} Z_m \right)} { (2l+1)! \; \; \Gamma \left( 2 - 2
\rm{i} Z_m
\right)}\nonumber \\
&& \times \int_{X_1 - \epsilon}^{X_2 + \epsilon} \frac{
 \left( \,^2 J_{lm} (\xi) \right)
 \left( \,^2 R_{lm}^{(I)} (\xi) \right) }{ \xi (\xi-1) }
\rm{d} \xi \, ,\quad\nonumber \\
&&  \,^2 R_{lm}^{(I)} \equiv \Bigg[1-\frac{1}{X}\Bigg]^{-iZ_m}\left[
\,^2 y_{{lm}}^{(I)} \right] \, .
\end{eqnarray}
The current source $\,^2 J_{lm} (r)$ is rather complicated:
\begin{eqnarray}
&& \hspace*{1.5pc}\,^2 J_{lm} (r) = \nonumber\\
 && \int_0^{2 \pi}\int_0^{\pi}\frac{ \left(r - \rm{i} a \cos \theta \right)^2 }{ \left( r_{_+} -r_{_-} \right)^2 }
 \rho^2 J_2 \left( \,_{-1} \bar{Y}_{lm} (\theta,\phi) \right)
 \sin \theta \rm{d} \theta \rm{d} \phi \; , \nonumber\\
 &&
\end{eqnarray}
\noindent where
\begin{eqnarray}
&& J_2 = \frac{-\Delta}{2 \sqrt{2} \rho^2 \left( r - \rm{i} a \cos
\theta \right)^2 }\nonumber \\
&&\times \left[ \sqrt{2} \left(
 \frac{\partial}{\partial r} - \frac{a}{\Delta} \frac{\partial}{\partial \phi} +
 \frac{1}{r - \rm{i} a \cos \theta} \right)
 \left( r - \rm{i} a \cos \theta \right) J_{\bar{m}} \right.
\nonumber \\
&& \hspace{-0.1 in}
 \left.
+ 2 \left(
 \frac{\partial}{\partial \theta} - \frac{\rm{i}}{\sin \theta}
 \frac{\partial}{\partial \phi} +
 \frac{\rm{i} a \sin \theta}{r - \rm{i} a \cos \theta} \right)
 \frac{\rho^2 \left( r - \rm{i} a \cos \theta \right)}{\Delta} J_n \right] \;
 ,\nonumber\\
 &&
\end{eqnarray}
and
\begin{eqnarray}
&& J_{\bar{m}} = \left[ \sqrt{2} (r - \rm{i} a \cos \theta)
\right]^{-1}
\nonumber \\[3pt]
&& \hspace{.1 in} \times \left[
 - \rm{i} a \cos \theta \tilde{J}^t
 - \rho^2 \tilde{J}^\theta
 + \rm{i} (r^2 + a^2) \sin \theta \tilde{J}^\phi
\right] \; ,
\\[3pt]
&& J_n = \frac{1}{2} \left[
   \frac{\Delta}{\rho^2} \tilde{J}^t
 + \tilde{J}^r
 - \frac{a \Delta}{\rho^2} \sin^2 \theta \tilde{J}^\phi
\right] \; .
\end{eqnarray}

\par A particularly relevant solution is for the external electromagnetic field of an
uncharged azimuthal current loop in the equatorial plane
\citep{bic76},

\begin{eqnarray}
&&b_{\ell m}=\frac{\delta_{m0} 4 \pi^{2}I^{\phi}(\ell +1)!\ell!}{\sqrt{2}(r_{+}-r_{-})(2\ell +1)!}\sqrt{\frac{\Delta(r_o)}{\rho(r_o)^{2}}}\nonumber\\
&&\times
\Bigg[\rm{i}\frac{r_{o}^{2}+a^{2}}{r_{+}-r_{-}}\,_{-1}\bar{Y}_{\ell
0} \left(\frac{\pi}{2},0\right)F(\ell +1,-\ell,1;X_{o})  \nonumber
\\
&&+\left(\rm{i}r_{o}\,_{-1}\bar{Y}_{\ell 0}\left(\frac{\pi}{2},0\right)-a\sqrt{\ell(\ell+1)}\,_{0}\bar{Y}_{\ell 0}\left(\frac{\pi}{2},0\right)\right) \nonumber \\
&&  \times  \left(X_{o}(X_o-1)F(\ell+2,1-\ell,2;X_{o})\right)\Bigg] \nonumber \\
&& E_{a}=E_{b}=0 \;.
\end{eqnarray}
where, $I^{\phi}$, is the azimuthal current in the current loop
evaluated in the ZAMO frames. Equation (A.19) determines the
solution at $r$ larger than the radial coordinate of the current
ring at $r_{o}$.
\par Equation (A.19) ignored the effects of charge and in general a
current ring in a rotating environment will have an induced charge
from a motional electromotive force. The motivation for segregating
the uncharged ring is that the charge contribution to the poloidal
magnetic flux is negligible in the calculation of Appendix C. If a
ring has a charge $q$ then there are two additional terms that need
to be added to Equation (A.19) for the source of the charged current
loop \citep{bic76}
\begin{eqnarray}
&&b_{\ell m}=\frac{\delta_{m0} 2 \pi q(l\ell+1)!\ell !}{\sqrt{2}(r_{+}-r_{-})^{3}(2\ell +1)!}\frac{\Delta(r_o)\rho(r_o)^{2}}{r_{o}g_{\phi\,\phi}(r_{o})}\nonumber\\
&&\times \Bigg[\rm{-i}ar_{o}^{2}(r_{+}-r_{-})\,_{-1}\bar{Y}_{\ell0}
\left(\frac{\pi}{2},0\right)F(\ell +1,-\ell,1;X_{o}) \nonumber
\\
&&+\Bigg(2\rm{i}Mar_{o}^{2}\,_{-1}\bar{Y}_{\ell0}\left(\frac{\pi}{2},0\right)-\big([r_{o}^{2}+a^{2}]^{2}-\Delta(r_{o})\rho(r_{o})^{2}\big)
 \nonumber \\
&& \times \sqrt{\ell
(\ell+1)}\,_{0}\bar{Y}_{\ell0}\left(\frac{\pi}{2},0\right)\Bigg)F(\ell+ 2,l-\ell,2;X_{o})\Bigg] \nonumber \\
&& E_{a}=E_{b}=\frac{1}{2}q \;.
\end{eqnarray}
The $E_{b}=\frac{1}{2}q$ term yields the Kerr Newman field of a
charged rotating black hole for a net charge near the black hole.
The $b_{lm}$ is the source from the azimuthal current produced by
the charge set into rotation by the frame dragging of spacetime.
Notice that the $b_{lm}$ term dies off like $\Delta(r_{o})$ for the
charged contribution to the external field in Equation (A.20) and
only dies off like $\sqrt{\Delta(r_{o})}$ for the uncharged current
ring in Equation (A.19).
\par The interior solution, Equation (A.7), for the uncharged current ring is given
by the source term \citep{bic76}
\begin{eqnarray}
&&a_{\ell m}=\frac{\delta_{m0} 4 \pi^{2}I^{\phi}(\ell +1)!\ell!}{\sqrt{2}(r_{+}-r_{-})(2\ell +1)!}\sqrt{\frac{\Delta(r_o)}{\rho(r_o)^{2}}}(-X_{o}^{-\ell})\nonumber\\
&&\hspace{-0.1in}\times
\Bigg[\rm{i}\frac{r_{o}^{2}+a^{2}}{r_{+}-r_{-}}\,_{-1}\bar{Y}_{\ell
0} \left(\frac{\pi}{2},0\right)\frac{\ell}{X_{o}}F(\ell +1,\ell +
1,2\ell + 2;X_{o}^{-1}) \nonumber
\\
&&+\left(\rm{i}r_{o}\,_{-1}\bar{Y}_{\ell 0}\left(\frac{\pi}{2},0\right)-a\sqrt{\ell(\ell+1)}\,_{0}\bar{Y}_{\ell 0}\left(\frac{\pi}{2},0\right)\right) \nonumber \\
&&  \times  \left(F(\ell,\ell+1,2\ell +2;X_{o}^{-1})\right)\Bigg] \nonumber \\
&& E_{a}=E_{b}=0 \;.
\end{eqnarray}

\section{Surface Currents in the ZAMO Frames}
Near the black hole, Equations (15) - (20) for the surface current
approximation to the thin flux tube needs to be formulated in a
general relativistic context. This will be done by integrating
Maxwell's equations in the ZAMO frame across the boundary of the
flux tube. One of the computational advantages of the ZAMO
orthonormal frame is that it is defined only up to a rotation in the
$\left( r, \theta \right)$ plane.  In the study of winds it is
useful to define a rotated ZAMO basis in which the unit vector
$\hat{e}_1$ is parallel to the poloidal component of the magnetic
field, $B^P$ \citep{pun08}. In terms of the Maxwell tensor in the
ZAMO frames,
\begin{equation}
B^1 \equiv B^P = F^{2\phi} \; , \quad \mbox{and} \quad B^2 = F^{\phi
1} = 0 \; .
\end{equation}
\noindent The basis\index{ZAMO frames!basis covectors} vectors in
the $\left( r, \theta \right)$ plane become
\begin{eqnarray}
&& \left[
 \begin{array}{c}
  \hat{e}_{1} \medskip \\
  \hat{e}_{2}
 \end{array}
\right] =
 \frac{1}{\left| B^P \right|}
\left[
 \begin{array}{cc}
  F^{\theta \phi} &
   F^{\phi r} \medskip \\
  -F^{\phi r} &
  F^{\theta \phi}
 \end{array}
\right] \left[
 \begin{array}{c}
  \hat{e}_r \medskip \\
  \hat{e}_\theta
 \end{array}
\right] \; .
\end{eqnarray}
Using $B^r = F^{\theta \phi}$ and $B^\theta = F^{\phi r}$, the basis
covectors in the rotated ZAMO frame are (note: $B^P =
\sqrt{\left(B^\theta \right) ^2 + \left( B^r \right) ^2}$ )
\begin{eqnarray}
&& \left[
 \begin{array}{c}
  \omega^1 \medskip \\
  \omega^2
 \end{array}
\right] =
 \frac{1}{\left| B^P \right|}
\left[
 \begin{array}{cc}
  B^r&
  B^\theta \medskip \\
 -B^\theta &
  B^r
 \end{array}
\right] \left[
 \begin{array}{c}
  \omega^r \medskip \\
  \omega^\theta
 \end{array} \right] \; .
\end{eqnarray}
This basis is more conducive to studying flux tubes. Partial
derivatives in the rotated ZAMO basis are found from (B.2) to be
\begin{eqnarray}
&& \frac{\partial}{\partial X^1} = \frac{B^r}{\left| B^P \right|}
\frac{\partial}{\partial X^r} - \frac{B^\theta}{\left| B^P \right|}
\frac{\partial}{\partial X^\theta} \; ,
\\
&& \frac{\partial}{\partial X^2} = \frac{B^\theta}{\left| B^P
\right|} \frac{\partial}{\partial X^r} + \frac{B^r}{\left| B^P
\right|} \frac{\partial}{\partial X^\theta} \; .
\end{eqnarray}
In the rotated ZAMO basis, the poloidal component of Ampere's Law is
found in \citet{pun08}:
\begin{eqnarray}
&\displaystyle
 \frac{\partial}{\partial X^0} F^{1 0} +
 \frac{1}{\alpha \sqrt{g_{\phi\phi}}}
 \frac{\partial}{\partial X^2}
 \left( \alpha \sqrt{g_{\phi\phi}} F^{12} \right) = \frac{4 \pi J^1}{c} \; .&
\end{eqnarray}
One can construct a local coordinate system that is momentarily at
rest with respect to the rotated ZAMO basis at any point in
spacetime $(r_{o}, \, \theta_{o})$,
\begin{eqnarray}
&&(X^{0}_{o},\, X^{1}_{o}, \,X^{2}_{o}, \, X^{\phi}_{o})\equiv
\nonumber\\
&& \hspace{-0.1 in}(X^{0}(r_{o}, \, \theta_{o}),\, X^{1}(r_{o}, \,
\theta_{o}), \,X^{2}(r_{o}, \, \theta_{o}), \, X^{\phi}(r_{o}, \,
\theta_{o}))\;,
\end{eqnarray}
where the last step means to evaluate the metric coefficients at
$(r_{o}, \, \theta_{o})$ and treat them as constants. Thus, the
coordinate system is orthonormal only at the origin. Then integrate
Equation (B.6) across the thin boundary layer at the edge of the
flux tube. Namely integrate over $-\epsilon <X^{2}_{o}<\epsilon$ and
take the limit of $\epsilon$ goes to zero,
\begin{eqnarray}
&&\frac{4\pi}{c}K^{P}_{\rm{Z}}(r_{\rm{in}}, \, \theta_{\rm{in}}) \approx -B^{\phi}(r_{\rm{in}}, \, \theta_{\rm{in}})\\
&&\frac{4\pi}{c}K^{P}_{\rm{Z}}(r_{\rm{out}}, \, \theta_{\rm{out}})
\approx B^{\phi}(r_{\rm{out}}, \, \theta_{\rm{out}})
\end{eqnarray}
The azimuthal component of Ampere's law does not simplify so nicely
in the rotated ZAMO basis. The ZAMO expression from \citet{pun08} is
\begin{eqnarray}
&&
 \frac{\partial}{\partial X^0} F^{\phi 0}
+\frac{1}{\alpha \sqrt{g_{\theta\theta}}}
 \frac{\partial}{\partial X^r}
 \left[ \alpha \sqrt{g_{\theta\theta}} F^{\phi r} \right]\nonumber
 \\
&& +\frac{1}{\alpha \sqrt{g_{rr}}}
 \frac{\partial}{\partial X^\theta}
 \left[ \alpha \sqrt{g_{rr}} F^{\phi \theta} \right]
\nonumber \\
&&
\quad+ 2 \left( \Gamma^{\phi} \smallskip_{0r} F^{0r} + \Gamma^{\phi}
\smallskip_{0\theta} F^{0\theta} \right) = \frac{4 \pi J^\phi}{c}
\; ,
\end{eqnarray}
where the connection coefficients are
\begin{eqnarray}
&& \Gamma^{\phi} \smallskip_{0r} = - \frac{ Ma \sin \theta } {
\rho^3 \left[ \left(r^2+a^2\right) \rho^2 + 2Mra^2 \sin^2 \theta
\right] } \nonumber \\
&& \times \left[ \left(r^2 - a^2\right) a^2 \cos^2 \theta + r^2
\left( 3r^2 + a^2 \right) \right]\\
 &&\Gamma^{\phi} \smallskip_{0\theta} = - \frac{ 2Mra^3 \sin^4 \theta \cos \theta }{ \rho^5 g_{\phi
\phi} } \left( r^2 + a^2 - 2Mr \right)^{1/2} \;.
\end{eqnarray}
Both connection coefficients are well behaved at the horizon. The
poloidal derivatives in Equation (B.10) can be rewritten in terms of
derivatives in the  rotated ZAMO basis using the inverse of
Equations (B.4) and (B.5). Then, integrate Equation (B.10) over
$-\epsilon <X^{2}_{o}<\epsilon$ and take the limit of $\epsilon$
goes to zero (with $X^{1}$ held fixed) to obtain
\begin{eqnarray}
&&\frac{4\pi}{c}K^{\phi}_{\rm{Z}}(r_{\rm{in}}, \, \theta_{\rm{in}}) \approx B^{P}(r_{\rm{in}}, \, \theta_{\rm{in}})\\
&&\frac{4\pi}{c}K^{\phi}_{\rm{Z}}(r_{\rm{out}}, \,
\theta_{\rm{out}}) \approx -B^{P}(r_{\rm{out}}, \,
\theta_{\rm{out}})
\end{eqnarray}

\section{Flux Dissipation in a Semi-Vacuum Magnetosphere} Without a pair creation mechanism
to quench the vacuum gap, at later times, the gap between the
ingoing and outgoing flows becomes larger and larger. The inner
current sources experience gravitational accretion toward the EH.
The outer current sources are still unbound and are driven off to
infinity by magneto-centrifugal forces in the flux tube. There is
nothing that will stop the semi-vacuum gap from spreading open.

\par Contrast this spreading gap with the familiar
configuration in a pulsar. Pulsar pair creation models (see Appendix
D) utilize the voltage drop across the magnetic field (see Equation
(D.2)) as the particle accelerating mechanism that initiates pair
cascades. Some of this voltage can be dropped along the length of
the gap, thereby accelerating particles in the gap. Is this a valid
mechanism for a weak accreted flux tube in isolation? The notion
that there is a voltage drop along the gap length is rooted in our
experience of the strong fields from super-conducting neutron stars
in pulsars. In Figure 3, if the field line angular velocity,
$\Omega_{F}$, is equal on both sides of the vacuum gap when the gap
starts to spread apart (and there is no reason it should not be,
since it was an instant before) then the voltage drop across the
field lines is equal above and below the gap by Equation (D.2).
Applying Faraday's law to a vacuum gap that starts to spread apart
there is a transient parallel poloidal electric field, $E^{P}$, in
the gap that is associated with the radiative decay of $B^{\phi}$ in
the gap. The poloidal electromagnetic field in the spreading gap is
composed of this transient displacement current and decaying
fringing poloidal magnetic fields. By Faraday's law, the sum of the
voltage drops around a closed poloidal loop near the center of the
gap will tend to zero in time. There is no residual voltage drop
along the field lines that increases as the gap spreads: and there
is no electromagnetic force that prevents the plasma below and above
the gap from moving off towards the horizon and infinity,
respectively.
 \par By contrast, in a pulsar, if the gap grows in time, the voltage drop in the gap
increases in time as well and so will the propensity for particle
acceleration in the gap. The fundamental difference is that the
sources of the gap magnetic field in the weak accreting flux tube
are not fixed in time and space. Conversely, the poloidal magnetic
field and the rotational EMF in the gap are persistent in the
pulsar; this dynamic is imposed by the star (not the magnetospheric
plasma as for the weak, accreting, magnetic flux tubes). Thus, one
cannot justify the use of the voltage drop in the gap as the source
of a pair cascade in the case of weak, accreting, isolated flux
tubes in the charge starved limit.

\par In the semi-vacuum region that forms between the sources, the
electromagnetic field transforms from MHD to radiative in nature.
The poloidal magnetic field topology changes as the current
disappears in the rapidly expanding vacuum gap. Initially, the
topology of the polodial field (in a flat spacetime analogy) in the
accreting thin flux tube resembles that of two coaxial solenoids
that extend to infinity in each hemisphere (see Equations (15) -
(20)). As the gap grows between the ingoing and outgoing current
sources, in each hemisphere of the EHM, the poloidal magnetic field
of the outgoing disconnected segment of the flux tube starts to
resemble that of two semi-infinite coaxial solenoids in each
hemisphere. The poloidal magnetic field of the ingoing segment near
the EH starts to resemble that of two short coaxial solenoids in
each hemisphere (see Figure C.1).
\par In this Appendix, the field configuration from
the inner portion of the severed flux tube on the background of a
semi-vacuum in the surrounding EHM is estimated. The charge is
considered so tenuous that it does not modify the vacuum fields from
the inner flux tube segment. The background charges move in response
to these fields, but are of insufficient quantity to create currents
strong enough to non-negligibly modify these fields. The entire
exact electromagnetic evolution is complicated. However, large
simplifications occur at late times as the outer flux tube moves far
out of the central vortex of the accretion flow and the inner
portion of the severed flux tube approaches the EH. This section
explores the large scale poloidal field from the inner severed flux
tubes. The analysis follows from the freezing of the flow and the
gravitational redshifting of axisymmetric current sources near the
event horizon that are quantified in the next section.

\begin{figure}
\includegraphics[width= 0.4\textwidth]{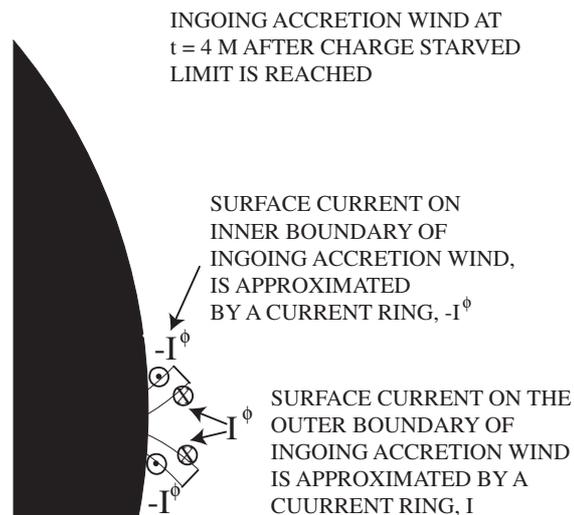}
\caption{ At $t\approx 4M$ (in geometrized units) after the time
snapshot in Figure 3, the outgoing severed flux tube segment has
been magneto-centrifugally slung out to large distances. The ingoing
portion of the severed flux tube has contracted toward the event
horizon. The particular case has $a/M=0.9$ and the top boundary of
the severed flux tube is at $r=1.08r_{+}$. The compact size suggests
replacing the azimuthal surface current source of $B^{P}$ in
Equations (21) and (22) with small azimuthal current rings. This
configuration defines the calculation that produces the field line
plot in the top frame of Figure C.2.}
\end{figure}
\subsection{The Effects of Gravitational Redshift on Maxwell's Equations}
\par Gravitational redshift results in the the freezing of the flow near the event horizon
\begin{eqnarray}
&&\frac{\rm{d}r}{\rm{d}t}=-\frac{\Delta}{r_{+}^{2} + a^{2}}\left[1+\rm{O}(\alpha_{\rm{Z}}^{2})\right] \; , \, r\gtrsim r_{+}\\
&&\lim_{t \rightarrow \infty} \left( r - r_{_+} \right) =
\mbox{constant} \times \rm{e}^{-2 \kappa t} \; ,
\end{eqnarray}
where $\kappa=\sqrt{M^{2}-a^{2}}/(r_{+}^{2} + a^{2})$ is the surface
gravity \citep{pun08}. Equation (C.2) indicates that as viewed from
asymptotic infinity all plasma approaches the black hole at a rate
that slows down exponentially, never actually reaching the event
horizon. In the remainder of this Appendix, the freezing of the flow
condition in Equation (C.1) will be used to estimate the time
evolution of the fields from the contracting severed flux tube.

\par  The quantity of interest for jet power from an EHM is
the large scale poloidal flux. The entire exact electromagnetic
evolution is complicated. There is electromagnetic radiation from
decaying fields and this radiation reflects off of the disk and the
centrifugal potential \citep{hod00,gle08}. However, after these
transients have decayed, and the current sources are near the event
horizon, the situation is simplified. The freezing of the flow and
the gravitational redshifting of the source term in Maxwell's
equations greatly simplifies the solution for the large scale
poloidal field from the severed flux tube at late times in the
semi-vacuum EHM. As first noted in \citet{pun89}, for axisymmetric
sources, one can implement the time stationary version of Maxwell's
equations in curved spacetime to achieve accurate solutions to the
large scale poloidal magnetic fields. To paraphrase the logic, as $r
\rightarrow r_{+}$, by the freezing of the flow all electromagnetic
sources seem to just corotate with the event horizon as seen by all
external observers. For axisymmetric sources there is no change
except for a slow poloidal advance toward the black hole given by
Equations (C.1) and (C.2). For axisymmetric sources, the time
dependent fields that are detected by external observers can be
approximated by treating the fields of the sources near the horizon
as changing adiabatically slowly and are approximately time
stationary. Thus, the time evolution of the large scale fields in
the semi-vacuum EHM arising from axisymmetric sources near the
horizon can be evaluated in terms of solutions to Laplace's
equations in Boyer-Lindquist. Equivalently, as $r\rightarrow r_{+}$,
$d\,t/d\, \tau \propto \alpha_{Z}^{-2}$, where $\tau$ is the proper
time \citep{pun08}. Thus, all time derivatives reflect the
gravitational redshift by being weighted by a factor $\propto
\alpha_{Z}^{2}$ relative to the time derivatives in the proper
frame. For axisymmetric sources of Maxell's equations, as
$r\rightarrow r_{+}$, this equates to an $\propto \alpha_{Z}^{2}$
contribution from displacement currents in Boyer-Lindquist
coordinates. Hence, time variability and displacement currents will
not affect the computation of the large scale poloidal magnetic
fields.

\par Appendix A is a review of the mathematical tools needed to study
Laplace's equations in the Kerr spacetime. The fundamental elements
of the field are the spin coefficients Equations, (A.8) - (A.10),
from which algebraic combinations produce the Faraday field strength
tensor in Boyer-lIndquist coordinates by means of Equations (A.1)
and (A.3). For external fields (those at larger radius than the
source), the source is expressed by the coefficients $b_{lm}$ that
are determined in (A.14). These calculations are in general rather
labor intensive. For simplicity, the axisymmetric approximations is
adopted. As an axisymmetric charge neutral source contracts toward
the black hole \citep{pun08}
\begin{equation}
\lim_{t \rightarrow \infty} b_{lm} (t) \sim \rm{e}^{- \kappa t} \; .
\end{equation}
The source term is gravitationally redshifted away and this result
is known as the "No-Hair Theorem" for charge neutral, axisymmetric,
electromagnetic sources.

\subsection{A Current Source Description of the Ingoing Severed Flux
Tube}
\par In the simulations of \citet{haw06,kro05}, the flow division
point in the EHM $\sim 1.4r_{+} - 1.5r_{+}$ for rapidly spinning
black holes. This will be taken as the starting location for the
vacuum gap in the following analysis. In Figures C.1 and C.2, the
black hole has $a/M=0.9$. From Equation (3), the EH is located at
$r_{+} =1.436 M$.

\par The inward spread of the vacuum gap is approximated by Equation (C.1) with the $O\left(\alpha_{Z}^{2}\right)$ correction set
equal to zero. This is a very crude approximation during the initial
spreading of the gap. Initially, $\alpha^{2}\approx 0.14$, so the
correction terms are not necessarily negligible. However, at late
times. as in Figure C.2, the top of the ingoing severed flux tube is
at $\alpha^{2} = 2.7 \times 10^{-2}$ in the top frame and
$\alpha^{2} = 3.2 \times 10^{-3}$ in the bottom frame. Thus, we
expect Equation (C.1), without correction terms, to be very accurate
in this part of the inflow. Even at the late times of interest,
there is an offset, $t_{o}$, to the infall times associated with the
error in using Equation (C.1), without correction terms, as the
equation of motion at the beginning of the inflow. Below the flow
division point, the electromagnetic force,
$\tilde{J}_{\nu}\tilde{F}^{r\, \nu}$ is directed inward
\citep{pun08}. The ingoing force means that the magnitude of the
ingoing radial velocity obtain from Equation (C.1), without
correction terms, is underestimated. Thus, the infall times computed
without the correction terms from the start of the gap spreading to
the configuration in the top frame of Figure C.2 are overestimated
and the offset, $t_{o} <0$. Since the infall time to reach the
configuration in the top frame of Figure C.2 computed from Equation
(C.1), without the correction terms, is 4M, and the total infall
time must be positive, the offset error is constrained to be
$t_{o}\sim M$. This does not affect the conclusions drawn from the
infall times in Section C.4 for two reasons. First, its magnitude is
relatively small and secondly we are using the infall times as a
maximum time for the flux tube to dissipate and decreasing this time
only strengthens the argument.

\par At $t\approx 4M$, after the vacuum gap starts to spread open, the
top boundary of the inner severed accretion flow will be at
$r\approx 1.08r_{+} =1.55M$ or by Equation (4),
$\alpha_{\rm{Z}}=0.163$. In Figures C.1 and C.2, it is evident that
the inner severed flux tube is very short and the
$K^{\phi}_{\rm{Z}}$ source for $B^{P}$ in Equations (21) and (22)
can be well approximated by azimuthal current loops. This
approximation does not include the affects of $K^{P}_{\rm{Z}}$,
which would source a local toroidal magnetic field near the loops
and inwards towards the horizon. It would also not accurately depict
fringing fields and fringing displacement currents near the severed
end of the flux tube. However, the interest here is the large scale
poloidal magnetic field. Based on the discussion of Section C.1, due
to the freezing of the flow in Equations (C.1) and (C.2), this
source model in Laplace's equations will produce representative
magnetic fields near the accretion disk and it will approximate the
background magnetic field in most of the semi-vacuum EHM
\citep{pun89}.. The current loops carry an equal and opposite
azimuthal current. However, the outer current loop at
$r=r_{\rm{out}}$ has a larger enclosed area than the current loop at
the inner boundary, $r=r_{\rm{in}}$, and therefore the magnitude of
the magnetic moment is larger (see the full general relativistic
calculational description in Equations (C.9) - (C.13)). Thus, there
is a net magnetic dipole moment.
\begin{figure*}
\begin{center}
\includegraphics[width= 0.9\textwidth]{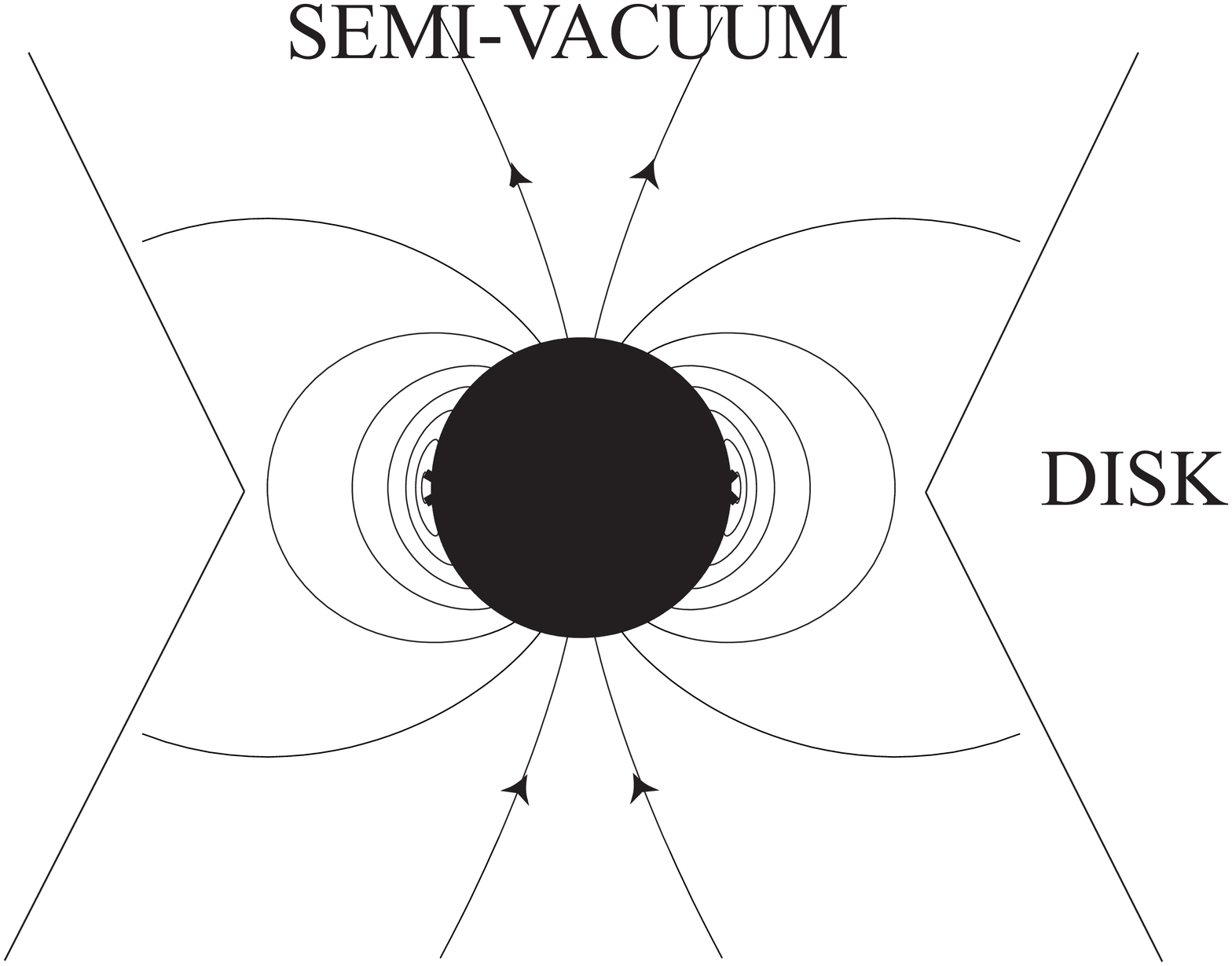}
\includegraphics[width= 0.9\textwidth]{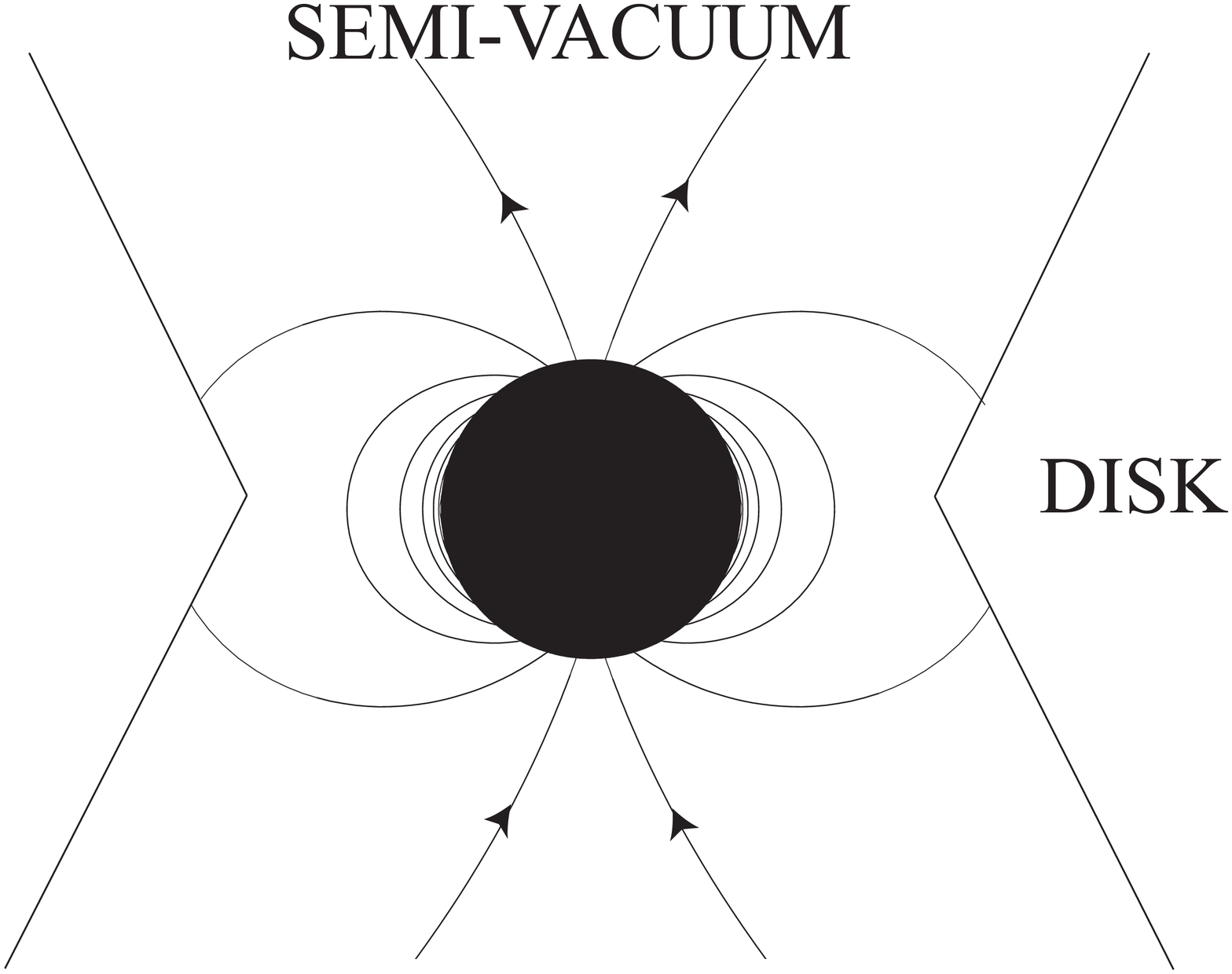}
\caption{ The ingoing portion of the severed flux tube has
contracted toward the event horizon to $r=1.08r_{+}$ in the top
frame at $t=4M$ after the vacuum gap starts to spread open. The
large scale poloidal magnetic field is plotted in Boyer-Lindquist
coordinates. The nature of the disk is very uncertain and it is
represented symbolically by a simple wedge. In the bottom frame, the
severed flux tube has contracted to $r=1.001r_{+}$ at $t=15M$ after
the vacuum gap starts to spread open. Note that the large scale
field is contracting towards the black hole. In the top (bottom)
frame the poloidal magnetic field in the disk is $\approx 2\%$
($\approx 0.4\%$) of the poloidal field strength in the original
flux tube when it entered the EHM. The bottom frame indicates that
poloidal magnetic flux is almost completely dissipated on a
timescale of one half of the rotational period of the ISCO.}
\end{center}
\end{figure*}
\par In order to solve for the global field of the four current loops in Figure C.1,
first note that $I^{\phi}$, the azimuthal current in the ZAMO frame,
used in the discussion of the current loop in the Appendix satisfies
from Equations (21) and (22),
\begin{equation}
I^{\phi}(r_{\rm{out}}) \approx - I^{\phi}(r_{\rm{in}})\;.
\end{equation}
Even though it is customary to use a ZAMO evaluated azimuthal
current in such calculations, \citet{chi75,bic76}, it is necessary
to understand if a well behaved ZAMO current provides any
pathological (unphysical) scalings near the event horizon. The
meaningful condition is that the globally defined, coordinate
independent, magnetic flux has no pathological behavior near the
event horizon. Implementing the rotated ZAMO basis (one poloidal
basis vector field is along the poloidal field direction and the
other poloidal basis vector field is orthogonal to the poloidal
field direction) from Equation (B.3), the flux in the flux tube is
\citep{pun08}
\begin{equation}
\Phi  = \int B^{P} \rm{d}X^{2} \, \rm{d}X^{\phi}=
(2\pi)\sqrt{g_{\phi \, \phi}}\int B^{P}\rm{d}X^{2}\;.
\end{equation}
The surface current in Equation (22) is related to $I^{\phi}$ by
\begin{equation}
K^{\phi}_{\rm{Z}}(r_{\rm{out}}, \, \theta_{\rm{out}}) = I^{\phi}
\delta [X^{2} -X^{2}(r_{\rm{out}}, \, \theta_{\rm{out}})]\;,
\end{equation}
where $X^{2}$ is a the poloidal coordinate orthogonal to the local
poloidal field direction defined in a coordinate system that is
instantaneously at rest with the ZAMO at $(r_{\rm{out}}, \,
\theta_{\rm{out}})$. Inserting this into Equations (22) and (C.5),
shows that the flux scales with $I^{\phi}$ and is only well behaved
if $I^{\phi}$ is well behaved. It is therefore concluded that the
ZAMO frame is suitable for evaluating the azimuthal current in this
study.
\par The inner flux tube at $t=4M$ in Figure C.1 is approximated by 4 current
loops that are located near the equatorial plane. Using Equation
(C.4), the current loop source term (A.19) evaluated at $r\gtrsim
r_{+}$ is approximated to find the net source term.
\begin{eqnarray}
&&b_{\ell m}\approx 2\frac{\delta_{m0} 4 \pi^{2}I^{\phi}(\ell+1)!\ell!}{\sqrt{2}(r_{+}-r_{-})(2\ell+1)!}\sqrt{\frac{\Delta(r_{\rm{out}})}{\rho(r_{\rm{out}})^{2}}}\nonumber\\
&&\times\left[\rm{i}\frac{r_{\rm{out}}^{2}+a^{2}}{r_{+}-r_{-}}\,_{-1}\bar{Y}_{\ell0}
\left(\frac{\pi}{2},0\right)F(\ell+1,-\ell,1;X_{\rm{out}})\right] \nonumber\\
&&- 2\frac{\delta_{m0} 4 \pi^{2}I^{\phi}(\ell+1)!\ell!}{\sqrt{2}(r_{+}-r_{-})(2\ell+1)!}\sqrt{\frac{\Delta(r_{\rm{in}})}{\rho(r_{\rm{in}})^{2}}}\nonumber\\
&&\times\left[\rm{i}\frac{r_{\rm{in}}^{2}+a^{2}}{r_{+}-r_{-}}\,_{-1}\bar{Y}_{\ell0}
\left(\frac{\pi}{2},0\right)F(\ell+1,-\ell,1;X_{\rm{in}})\right]\;,
\end{eqnarray}
where by Equation(A.4), $X=1$ at the event horizon, thus making two
of the terms that would have arisen from (A.19) negligible. The
neglect of these terms means that the fringing fields near the
horizon are not accurately depicted. This greatly simplifies the
algebra and does not affect the global poloidal magnetic field
calculation in the EHM which is the quantity of interest.
\footnote{These smaller terms are included in the final field plot
in Figure C.2. This is done to make the field solution vary smoothly
near the EH, but it should be remembered that due to the numerous
approximations in this discussion that the field configuration near
the EH is not accurate}

Further simplification at $r\gtrsim r_{+}$ yields

\begin{eqnarray}
&&b_{\ell m}\approx 2\frac{\delta_{m0} 4
\pi^{2}I^{\phi}(\ell+1)!\ell!}{\sqrt{2}(r_{+}-r_{-})(2l+1)!}\nonumber\\
&& \hspace{-0.2in}\times \left[\frac{\sqrt{\Delta(r_{\rm{out}})}(r_{\rm{out}}^{2}+a^{2})}{r_{\rm{out}}}-\frac{\sqrt{\Delta(r_{\rm{in}})}(r_{\rm{in}}^{2}+a^{2})}{r_{\rm{in}}}\right]\nonumber\\
&&\hspace{-0.2in}\times\left[\rm{i}\frac{1}{r_{+}-r_{-}}\,_{-1}\bar{Y}_{\ell0}
\left(\frac{\pi}{2},0\right)F(\ell+1,-\ell,1;0.5(X_{\rm{out}}+X_{\rm{in}}))\right] \nonumber\\
\end{eqnarray}
\subsection{The Large Scale Poloidal Magnetic Field of the Ingoing
Severed Flux Tube}

The resultant external large scale poloidal magnetic field is
derived by inserting Equation (C.8) into Equations (A8) - (A12) to
solve for the spin coefficients. Then using (A.1) and (A.3) the spin
coefficients determine the Boyer-Lindquist evaluated fields.

The result is plotted in Figure C.2. There is an intersection of a
weak field line with the disk that is not indicated since there is
an uncertain interaction. The field strength at the disk in the top
frame is $<2\%$ of the original weak field strength of the accreted
flux tube, thus no significant interacting is expected. There is an
important simplification of the large scale poloidal field when the
ingoing severed flux tube is located near the EH due to the term in
the numerator of Equation (C.8),
\begin{equation}
\mathcal{M}\equiv\frac{\sqrt{\Delta(r_{\rm{out}})}(r_{\rm{out}}^{2}+a^{2})}{r_{\rm{out}}}-\frac{\sqrt{\Delta(r_{\rm{in}})}(r_{\rm{in}}^{2}+a^{2})}{r_{\rm{in}}}\;.
\end{equation}
>From Equations (4) and (6) this corresponds to a difference in lapse
function. The lapse function is significantly smaller (the
gravitational redshift is significantly larger) for the inner
current loops than the outer current loops. For example, in Figure
C.1, the inner current loops are located at $\alpha_{\rm{Z}}= 0.08$
and the outer current loops are located at $\alpha_{\rm{Z}}= 0.14$.
Thus, the outer current loops contributes almost two times as much
to the large scale poloidal field as the inner loops. This effect
becomes more pronounced as the severed inner flux tube (current
loops in the approximation) contracts closer to the event horizon.
The current loop configuration in Figure C.1 is used as a source of
Laplace's equations to compute the magnetic field configuration in
the top frame of Figure C.2. The resultant field line plots are
computed by the same methods employed by \citet{bic80}. Field line
plots have been traditionally used to show the behavior of the large
poloidal magnetic field as a current source is moved closer to a
black hole and are well suited for our purposes. The particular
choice of current loops in Figure C.1 at $r=1.02r_{+}$ and $r=1.06
r_{+}$ actually produces a resultant source term in Equation (C.8)
that is equivalent to a single current loop at $r=1.05 r_{+}$ in the
equatorial plane. The field line plot was produced by Tomas Ledvinka
and generously supplied to \citet{pun08} as Figure 4.9.
\par The bottom frame of Figure C.2 is the same calculation with the
poloidal field of the severed flux tube analyzed at $t\approx 15M$
after the vacuum gap starts to spread. The top boundary of the
severed flux tube is now at $r=1.001r_{+}$ or
$\alpha_{\rm{Z}}=0.057$. The poloidal magnetic field strength at the
disk is $\approx 0.37\%$ of the original field strength of the
accreted flux tube. Comparing the top frame and bottom frames of
Figure C.2 illustrates the No-Hair Theorem for axisymmetric charge
neutral sources in the Kerr spacetime for this particular example.
Comparing the top frame in Figure C.2 to the bottom frame of Figure
C.2 shows that the poloidal magnetic field is being contracted into
a circular set of O-points where the magnetic flux is dissipated.
Furthermore. this circular set of O-points is slowly contracting
toward the event horizon. This is manifested on global scales by the
factor of 5 reduction of the poloidal field strength at the
accretion disk in the bottom frame. The global poloidal magnetic
field from the ingoing portion of the severed flux is redshifted
away. The point of the approximate calculation is not to find the
exact field at a given time and point in space. It is to show that
the effects of gravitational redshift are overwhelming and the inner
accreting severed flux tube will quickly have its large scale fields
redshifted away.
\par Express
the Boyer-Lindquist magnetic field plotted in the top frame of
Figure C.2 in the orthonormal ZAMO frame at large r
\begin{eqnarray}
&&\lim_{r \rightarrow \infty} B^{\theta} = \frac{2\pi
I^{\phi}\mathcal{M}(\sin{\theta})}{r^{3}}\; ,\\
&&\lim_{r \rightarrow \infty} B^{r} = \frac{2\pi I^{\phi}\mathcal{M}
(2\cos{\theta})}{r^{3}}\; ,
\end{eqnarray}
where $\mathcal{M}$ was defined in Equation (C.9). These are the
poloidal magnetic field components of a magnetic dipole with a net
magnetic moment is $2\pi\mathcal{M}I^{\phi}$. The net magnetic
dipole moment is the sum of the two larger dipole moments from the
outer current loops, each with a magnetic moment, $m_{\rm{out}}$,
\begin{equation}
m_{\rm{out}}=\pi
I^{\phi}\frac{\sqrt{\Delta(r_{\rm{out}})}(r_{\rm{out}}^{2}+a^{2})}{r_{\rm{out}}}\;,
\end{equation}
minus the sum of the two smaller dipole moments from the inner
current loops, each with a magnetic moment, $m_{\rm{in}}$,
\begin{equation}
m_{\rm{in}}=\pi
I^{\phi}\frac{\sqrt{\Delta(r_{\rm{in}})}(r_{\rm{in}}^{2}+a^{2})}{r_{\rm{in}}}\;.
\end{equation}
Since $\Delta$ tends to zero as $r \rightarrow r_{+}$, $m_{\rm{in}}$
attains a significantly smaller value than $m_{\rm{out}}$, more so
than would be expected from a pure flat space analogy as a
consequence of gravitational redshifting.
\par In order to make this approximation scheme for the complicated calculation complete, a discussion of the
effects of rotational induced charge on the solution must be
included. Since this is a rotating system there will be a
rotationally induced cross field potential drop and associated
electric field that is orthogonal to the poloidal magnetic field
direction. In the rotated ZAMO basis of Appendix B, this electric
field has a very simple form \citep{pun08}
\begin{equation}
 F^{2\,0} =\beta_{F}^{\phi} B^{P}\;.\,
\end{equation}
where $\beta_{F}^{\phi}$ is the angular velocity of the local field
as measured by the local ZAMO. By Gauss's law there msut be a
surface charge density on the walls of the flux tube. The field
lines begin on one wall of the flux tube and end on the opposite
wall. The net charge on one flux tube wall, $q$, is equal and
opposite that, $-q$ on the other wall in order to source and sink
the field lines. Now consider the contributions due to a net charge
on the current loop in the ZAMO frame as shown in Equation (A.20).
There are two terms, in addition to Equation (A.19), that contribute
to the large scale poloidal magnetic field. The dominant term is the
Kerr-Newman field term from $E_{b}$. For a charge near the EH, this
term creates the electromagnetic field of a Kerr-Newman (charged,
rotating) black hole. If one computes Gauss's law on a closed
surface just outside the severed flux tubes in Figure C.1, then the
net rotationally induce charge on the flux tubes will identically
cancel. Thus, for the total configuration, the total charge is zero
and $E_{b}=0$ \citep{bic76}. Thus, there is no large scale
Kerr-Newman poloidal magnetic field.

There is another term in Equation (A.20) that is from the rotational
motion of the charge and the current that it produces
\begin{eqnarray}
&&\lim_{r \rightarrow r_{+}} b_{lm}(\rm{rotating \; \rm{charge}})
\propto q\alpha_{Z}^{2}.
\end{eqnarray}
By comparison from Equations (A.19), (C.5) and (C.6), the charge
neutral current source
\begin{eqnarray}
&&\lim_{r \rightarrow r_{+}} b_{lm}(\rm{charge \; neutral \;
\rm{current}}) \propto \Phi \alpha_{Z}.
\end{eqnarray}
Thus, combining Equations (C.15) and (C.16)
\begin{eqnarray}
&&\lim_{r \rightarrow r_{+}}
\frac{b_{lm}(\rm{rotating}\;\rm{charge})}{ b_{lm}(\rm{charge \;
neutral\; \rm{current}})} \propto \frac{q}{\Phi}\alpha_{Z}
\rightarrow 0.
\end{eqnarray}
Since $q$ and $\Phi$ are well behaved and frame independent, the
charge neutral term will dominate near the event horizon. This
approximation was used in the computation of the field lines in
Figure C.2.
\par With the approximation of ignoring the charge contribution to
the large scale poloidal flux, the field line configuration is
completed in Figure C.2, by choosing a compatible source term for
the interior solution for $r_{+}<r<r_{o}$, $a_{lm}$. This interior
solution is that of the uncharged current rings only in Equation
(A.21). Even though this creates continuous poloidal magnetic field
lines, it should be noted that this simplification does not yield an
accurate depiction of the interior solution near the event horizon.
However, as has been repeatedly emphasized, this complicated detail
is not relevant to the study of the large scale poloidal magnetic
flux.
\subsection{Results in the Context of Global Accretion}
\par Another item to consider is the time scale for a flux tube to
dissipate, $t_{\rm{dis}}$, compared to the dynamical time scale,
$t_{\rm{dyn}}$, in which a new weak patch of poloidal flux is fed
into the EHM. Based on the physical assumptions 2) and 3) expressed
in Section 2.2, this analysis does not consider a constant flood of
magnetic flux as the physical state of flux deposition. Flux
deposition occurs over long periods of time and since it is not a
constant flood of magnetic field, but the occasional deposition of a
flux tube, $t_{\rm{dyn}}$ would be longer than the inflow time of
the disk plasma. The shortest possible dynamical time scale of the
accretion flow is the orbital period of the innermost stable orbit
(ISCO). In Boyer-Linquist coordinates, there is a slow radial inward
drift near the ISCO, but the plasma motion is mainly rotational as
it completes many revolutions as it approaches the ISCO
\citep{sad11}. The Keplerian angular velocity as viewed in the
stationary frames at asymptotic infinity is
\begin{equation}
\Omega_{\rm{kep}}(r) = \frac{M^{0.5}}{r^{1.5}+aM^{0.5}}\;.
\end{equation}
For $a/M=0.9$, the ISCO is at $r=2.32 M$, corresponding to an
orbital period of $T=28M$. Thus, from Figures C.1 and C.2, the flux
tube dissipates in less than half an orbital period of the ISCO.
Thus based on assumptions 2) and 3) any reasonable dynamical time
for flux deposition would satisfy

\begin{equation}
t_{\rm{dis}} \ll t_{\rm{dyn}}\;.
\end{equation}
This does not mean that the magnetic flux is identically zero in the
EHM. Small amounts of flux could exist (especially episodically)
based on the dissipation rate and inflow rate of magnetic flux
balance even in the absence of efficient pair creation. The main
idea is that since the flux dissipates on time scales that are much
less than any dynamical scale of the accretion disk, in the charge
starved limit, flux will not build up in the EHM. Old flux will
dissipate faster than new flux is deposited. The EHM is a sink for
flux not a reservoir. Thus, flux never builds up to the point that a
pair creation scenario would be effective (as discussed in Appendix
D). This also does not mean that mathematically based models in
which one floods the EHM with strong poloidal field would not
sustain itself through pair creation. However, it is unclear if that
assumed dynamic is relevant to low luminosity accretion systems.
\par Finally, it is noted that the time evolution does not support the idea that the
flux tube will induce a surface current on the inner disk that will
in turn be the source of the flux tube and maintain the flux. The
tangential magnetic field does induce a surface current on the inner
wall of the accretion disk. However, this field is transient and
decays as the extent of the severed region grows (see Figure C.2).
Furthermore, based on Figure C.2, the sign of the surface current is
the opposite of what is required to maintain the flux. In summary,
surface currents in the disk do not prevent accreted, thin, isolated
flux tubes in a charge starved EHM from dissociating. Accreted weak
flux is dissipated in the EHM, not accumulated.

\subsection{Summary of the Results of the Calculation}

In this Appendix, we calculated multiple items. Some were required
to established the relevant approximations (these can be useful for
future calculations performed by other researchers as well) and some
relate directly to the dynamics of the weak accreted flux tubes in
the EHM. The last items are the most important to this research and
we list these first.
\begin{enumerate}
\item Using a very elaborate spin coefficient formalism that
is described in the Appendix A, it is shown that coaxial helical
current flows that contract toward the event horizon in each
hemisphere will produce a large scale poloidal magnetic field that
decays to very small values as the horizon is approached. This is
shown graphically in Figure C.2. Quantitatively, when the top end of
the accreted coaxial solenoids is at $\sim 1.08r_{+}$ ($\sim
1.001r_{+}$) the field at the accretion disk is $\approx 2\%$
($\approx 0.37\%$) of the original field strength in the accreted
flux tube.
\item In Section C.4, it is demonstrated that time
scale for the fields to decay in the EHM is much shorter than any
plausible dynamical time scale in the accretion flow and flux will
dissipate faster than it will accumulate. This prevents the growth
of a significant EHM.
\item In Figure C.2 and Section C.4, it is demonstrated that as
a charge starved flux tube dissipates in the EHM, it does not induce
a current on the inner wall of the accretion disk that can sutain
the flux within the EHM.
\item In Section C.2, Equations (C.5) and (C.6), it is shown that the ZAMO evaluated azimuthal current
used by other researchers for similar calculations is an appropriate
current to consider for studying the global changes in the poloidal
magnetic field as sources approach the event horizon.
\item It is shown in Section C.3, Equation (C.17), that the charge (Kerr-Newman) contribution to
the contracting current loop configuration is lower order and can be
neglected near the event horizon.
\item In Section C.3, Equations (C.12) and (C.13), it is shown that the magnetic moment of an
azimuthal current source is reduced by gravitational redshift near
the event horizon.
\end{enumerate}
The first point is critical for our understanding of the dynamics of
weak accreted flux tubes in the EHM in the charge starved limit. In
a charge starved EHM, without plasm injection, the large scale
current system supporting an isolated accreted flux tube severs at
the flow division point. In the lower segment, the plasma and its
currents will then accrete toward the horizon. The flux tube near
the horizon is not aware of the severing due to causality. As long
as charges are being supplied from upstream, the lower portions of
the flux tube are unaffected. The top end of the flux tube is
therefore treated as a contact discontinuity that signals the change
to the lower section of the flux tube. The ingoing flow from the
charge depleted region below the vacuum gap makes the inner severed
flux tube appear like two contracting coaxial solenoids in each
hemisphere. Thus, the calculation presented in this section
indicates that at late times. the large scale poloidal magnetic
field of the charge depleted inner flux tube will approximately be
an ever weakening dipolar magnetic field.

\section{Vacuum Gap Induced Pair Cascades on Weak Accreted Flux
Tubes} This appendix discusses standard pair creation models in the
EHM with two additional considerations that are germane to the new
model of the EHM. First, the notion that the magnetosphere is not
fully developed is incorporated and one is considering the
deposition of isolated weak flux tubes in the EHM. Secondly, the
implications of the very low luminosity of the accretion flow in M87
is studied. In any vacuum gap pair creation model, if the magnetic
field in each accreted flux tube is sufficiently weak, the electric
field in the gap is incapable of producing enough pairs to seed the
currents and prevent further gap growth \citep{bla77}. This is in
contrast to rapidly rotating neutron stars (NSs) with strong
magnetic fields in pulsars. It is more analogous to weaker field NSs
which cannot support pulsars that are below (far below in this case)
the pulsar death line \citep{che93,har02}. The electric field that
accelerates particles to high energy and initiates a pair cascade
ultimately arises from the electromotive force induced by the
rotation of the magnetic field. This is proportional to the poloidal
magnetic field strength, $B$. Thus, regardless of the pair creation
mechanism, there exists a $B$ small enough that the threshold
electric field for pair production is not attained. For any pair
creation scenario envisioned, one can assume that $B$ for the
accreted flux tubes is below this threshold. However, before making
this assumption, pair creation models are considered in detail in
the EHM. In spite of the detailed discussion to follow, it is
important to remember that the assumption is the accretion of weak
magnetic flux with field strengths lying far below the effective
pair creation death line. Furthermore, one should not lose sight of
the fact that the field strength of accreted flux is unknown in the
example of M87 and estimates depend on many assumptions
\citep{kin14,kin15,had16}.
\par As noted in the references in the Introduction, many pair cascade models
based on vacuum gaps have been posited in the literature. They
depend on different assumptions and \citet{hir16} provide an in
depth comparison and contrast of a few of these. In order to make
the statements concrete these models are referenced to the context
of M87.

\subsection{Pair Cascades Based on Inverse Compton
Scattering}
 A possible model of pair creation is one in which inverse Compton
(IC) scattering of the soft photon field by seed electrons
accelerated in the semi-vacuum gap can initiate pair creation. It is
a two step process.
\begin{enumerate}
\item First, the high energy electrons up-scatter the soft
photons to gamma ray energies.
\item Next, these high energy gamma rays
scatter off the soft photon field, producing the electron-positron
pairs that fill the magnetosphere.
\end{enumerate}
These models were motivated by radiatively efficient accretion
scenarios in which there is a strong soft photon field. However,
there is no detection of the soft photon field in the example of
M87. The broadband spectrum based on high spatial resolution
observations is consistent with that of a synchrotron emitting jet
\citep{why04,lei09,pri16}. Thus, observations can not provide an
estimate of the soft photon flux from the inner accretion flow that
would support a putative IC based pair cascade. It is pointed out in
the setup of the IC based cascade model of \citet{bro15} that it is
``unclear" what the soft photon flux would be in M87 and that the
number density produced by the cascade strongly depends on the soft
photon flux. Furthermore, an IC based pair cascade ``gap solution
ceases to exist" if the soft photon flux is sufficiently small
\citep{hir16}. Thus, transferring the IC based gap models to
inefficient accretion flow environments, such as that in M87, is not
straightforward.  It is assumed that the soft photon flux is too
small for the process to operate and the following analysis
concentrates on models that assume that curvature radiation is the
source of the high energy gamma rays that initiate pair production.
This does not have the ambiguity of invoking an undetected soft
photon flux.

\subsection{Curvature Radiation Based Gap Models}

This section is a review of curvature pair production processes in
the context of EHM formation by the accretion of isolated, weak,
strands of flux as opposed to pair production in a fully developed
magnetosphere. One curvature radiation based gap model is introduced
for the purpose of an example only \citep{che93}. This pulsar gap
model has been adapted to Kerr-Newman black holes \citep{pun98}.
Even though other pair creation models have proven to have better
predictive power for $\gamma$-ray emission than \citet{che93}, such
as \citet{hig97}, it is beyond the burden and scope of this paper to
compare and contrast the different gap models. This is already
actively debated in the literature (see the references in the
Introduction). This burden is circumvented by assuming that the
field strength of an accreting, isolated, thin flux tube is weaker
than that which is required to initiate a cascade. Note, there is no
claim that the \citet{che93} method is the best model of a gap for
M87. It is chosen because it is straightforward to compute
analytically.

The only required ingredients of the Chen and Ruderman model are
rotation and a magnetic field that is not purely radial. This
process does not depend on the unknown weak soft photon flux. In
this model, it is the rotationally induced voltage drop across the
magnetic field lines, $\Delta V$, that is the energy source for the
pair cascade. The condition for pair creation is
\begin{equation}
\left(\frac{e \Delta V}{m_{e} c^{2}}\right)^{3}
\frac{\hbar}{2m_{e}cr_{c}}\frac{h}{r_{c}}\frac{B}{B_{g}}  >
\frac{1}{15} \; ,
\end{equation}
where $m_{e}$ is the mass of the electron, $r_{c}$ is the radius of
curvature of the magnetic field, $h$ is the gap height, $B_{g} =4.4
\times 10^{13}$ G is the critical magnetic field strength and $B$ is
the EHM poloidal magnetic field near the EH. The voltage drop across
a thin flux tube, rotating with a field line angular velocity,
$\Omega_{F}$, and a total flux $\delta \Phi$ is \citep{pun08}
\begin{equation}
\Delta V\approx \frac{\Omega_{F}\delta \Phi}{2 \pi c} \;.
\end{equation}
Consider a flux tube that threads a fraction, ${\mathcal F}$, of the
total surface area of a hemisphere of the event horizon. From the
expression for the hemispheric surface area of the EH,
$SA_{\rm{EH}}$, in \citet{pun08}, one can write the more convenient
expression,
\begin{equation}
\delta \Phi \equiv {\mathcal F}SA_{\rm{EH}}= {\mathcal F}2\pi
(r_{+}^{2} + a^{2})B \;,
\end{equation}
where the radius of the event horizon is $r_{+} = M + \sqrt{M^{2} -
a^{2}}$.  The analog of ${\mathcal F}$ in the standard pulsar model
is the fraction of the neutron star surface area, near the polar
cap, that comprises the open field lines of the magnetosphere
\citep{che93}. A field line angular velocity intermediate between
the seminal value of \citet{bla77} and the numerical values of
\citet{mck12} is chosen,
\begin{equation} \Omega_{F} = 0.4 \Omega _{H}=
\frac{a}{5Mr_{+}}\;,
\end{equation}
where $\Omega_{H}=a/(2Mr_{+})$ is the EH angular velocity as viewed
by stationary observers at asymptotic infinity.

For illustrative purposes, one can use the simulated field lines
such as those from \citet{bec09} in Figure 1 in order to estimate
$r_{c}$. The minimum value occurs juxtaposed to the EH, $r_{c}
\gtrsim 8r_{+}$. The value approximates the flux tube out to
$\approx 2.5r_{+}$. Thus, $r_{c} \approx 8r_{+}$ is chosen. To be
quantitative, this analysis is applied to M87, thus all numbers are
based on the value of M for M87. A value of $h\approx 0.5 r_{+}$ is
also chosen. The gap height choice is similar in terms of the radius
of the compact object to that obtained in a pulsar dipolar field
\citep{che93}. Choosing $r_{c}$ and $h$ is somewhat subjective.
Beyond $\sim 5 r_{+}$ of the EH, $r_{c}>>r_{+}$. Thus, if the gap
grows large, $(h/(r_{c})^2)$ will decrease and the pair creation
threshold value of B must increase to compensate in equation (D.1).
A value of ${\mathcal F}=0.1$ is chosen (see Figure 2). Based on
these pulsar models, Equations (D.1)-(D.4) imply a death-line
(insufficient electromotive force to drive pair creation) if $B\sim
225 $ G ($B\sim 1440 $ G) for $a/M = 0.99$ ($a/M \sim 0.1$).
\par The original analysis in \citet{bla77} for
the curvature radiation process, Equation (2.8), is similar to the
analysis for pulsars \citep{che93}. It is consistent with the
minimum B-field analysis above although they obtain a smaller number
due to different assumptions. They consider a fully established
$2\pi$ steradian hemispheric magnetosphere as an initial state as
opposed to a thin flux tube, their characteristic angular frequency
is 5 times the value in equation (D.4) and the radius of curvature
is simply $M$. Making these adjustments reconciles the two
estimates.

\subsection{Discussion}
The vacuum gap analysis considered here is consistent with the
analysis of the assumptions provided in the seminal model of
\citet{bla77} in which it was stated upfront that ``If the field
strength is large enough, the vacuum is unstable to a cascade
production of electron-positron pairs and a surrounding force-free
magnetosphere will be established". In other words, there is a
minimum field strength for the cascade models to work.

\par The pulsar-based analysis in Section C.2 is for demonstration
purposes only. Other gap models can produce different minimum field
strengths. The example demonstrates the fact that any pair creation
scenario must rely on estimating the parametric details. This
introduces a significant uncertainty in any method. Based on the
crude guide provided by the pulsar-based estimates, it is clear that
there is some field strength (which, given the uncertainties, could
be below the limit derived from this calculation) which will be too
low to allow an accreted flux tube to sustain itself through pair
production in the vacuum gap. The new EHM model consider the
dynamics if the accreted flux tubes have field strengths this low.

\end{appendix}
\end{document}